\documentclass[a4paper,11pt]{article}
\pdfoutput=1 

\usepackage{jcappub,aas_macros} 

\usepackage[T1]{fontenc} 

\newcommand{\nm}[0]{n_{\rm{max}}}
\newcommand{\M}[0]{\mathrm{Mpc}/h}
\newcommand{\iM}[0]{h/\mathrm{Mpc}}
\usepackage{hyperref}
\usepackage{fontawesome} 

\title{\boldmath An Analytic Hybrid Halo + Perturbation Theory Model for Small-scale Correlators: Baryons, Halos, and Galaxies}

\author[a,b,1]{James M. Sullivan,\note{Corresponding author.}}
\author[a,b,c]{Uro\v{s} Seljak}
\author[d,b]{Sukhdeep Singh}

\affiliation[a]{Department of Astronomy, University of California, Berkeley, CA 94720, USA}
\affiliation[b]{Berkeley Center for Cosmological Physics, University of California, Berkeley, CA 94720, USA}
\affiliation[c]{Lawrence Berkeley National Laboratory, One Cyclotron Road, Berkeley, CA 94720, USA}
\affiliation[d]{McWilliams Center for Cosmology, Carnegie Mellon University, Pittsburgh, PA 15213, USA}

\emailAdd{jmsullivan@berkeley.edu}
\emailAdd{useljak@berkeley.edu}
\emailAdd{sukhdeep@cmu.edu}

\abstract{
We update Halo Zeldovich Perturbation Theory (HZPT, \cite{2015PhRvD..91l3516S}), an analytic model for the two-point statistics of dark matter, to describe halo and galaxy clustering, and galaxy-matter cross-correlation on nonlinear scales.
The model correcting Zeldovich has an analytic Fourier transform, and therefore is valid in both configuration space and Fourier space.
The model is accurate at the $2\%$-level or less for $P_{mm}~(k< 1~\iM),~ P_{hm}~(k< 1~\iM),~P_{hh}~(k< 2~\iM),~ P_{gm}~(k< 1~\iM),~ P_{gg}~(k< 1~\iM),~ \xi_{mm}~(r> 1~\M),~ \xi_{hm}~(r> 2~\M),~ \xi_{hh}~(r> 2~\M)$, \ $\xi_{gm}~(r> 1~\M)$, $\xi_{gg}~(r> 2~\M)$, for LRG-like mock galaxies.
We show that the HZPT model for matter correlators can account for the effects of a wide range of baryonic feedback models and provide two extended dark matter models which are of $1\% ~(3\%)$ accuracy for $k < 10~ (8) ~\iM$.
We explicitly model the non-perturbative features of halo exclusion for the halo-halo and galaxy-galaxy correlators, as well as the presence of satellites for galaxy-matter and galaxy-galaxy correlation functions.
We perform density estimation using N-body simulations and a wide range of HOD galaxy mocks to obtain correlations of model parameters with the cosmological parameters $\Omega_{m}$ and $\sigma_{8}$. 
HZPT can provide a fast, interpretable, and analytic model for combined-probe analyses of redshift surveys using scales well into the non-linear regime.
}

\begin{document}
\maketitle
\flushbottom

\section{Introduction}
\label{sec:intro}

The goal of large-scale structure (LSS) analysis is to extract cosmological information from the nonlinear matter density field.
Nearly all modern cosmological analyses are built upon two-point statistics that probe this field \cite{2017MNRAS.470.2617A,2018PhRvD..98d3526A}.
Increasingly precise measurements of two-point correlators in modern galaxy surveys demand percent-level accuracy of theoretical models of these correlators.
This is especially true on smaller scales where the effect of survey sample variance is minimal and non-perturbative effects dominate.
However, models of two-point statistics require a tradeoff between the scales they access and the amount of theoretical control they have.

Perturbation theory (PT) provides an analytic model of the density field on large scales \cite{2002PhR...367....1B}.
Two-point correlators in perturbative models are limited by the scale at which nonlinear effects dominate the dark matter dynamics
which is a much 
larger scale than the minimum scale to which current surveys are sensitive \cite{2018PhR...733....1D}. 
However, PT models remain attractive due to the control over theoretical errors they afford in their domain of validity. 
Extensions of perturbation theory based on an effective fluid description of the density field (EFT) have pushed deeper into the quasi-linear regime \cite{2012JHEP...09..082C,2012JCAP...07..051B,2015JCAP...09..014V,2014JCAP...05..022P,2016JCAP...03..057V}. 
Such extended perturbation theory models have recently been used successfully for analysis of cosmological parameters, though with nuisance parameters that are fitted to numerical simulations (e.g. \cite{2020JCAP...05..042I,2020JCAP...05..005D,2020JCAP...06..001C}).
However, there is a limit to any perturbative model, even in the EFT framework, as non-perturbative effects and nonlinear gravitational evolution dictate the behavior of the density field on scales less than a few $\rm{Mpc}/h$.
In fact, it is clear that perturbation theory does not actually converge to the fully nonlinear result on smaller scales, at least in one dimension, at infinite order \cite{2016JCAP...01..043M,2018JCAP...05..039P}. 
This is due to the fundamentally non-perturbative nature of small-scale dark matter dynamics. 

An alternative analytic model that includes non-perturbative effects in the form of halo formation is the 
halo model \cite{2000MNRAS.318..203S,2000MNRAS.318.1144P,2002PhR...372....1C,2000ApJ...543..503M}.
The halo model assumes that all dark matter is tied up in gravitationally bound, non-overlapping halos, which have a prescribed density profile and an abundance set by the halo mass function.
Two-point matter correlators are computed by way of mean halo profiles integrated over the halo mass function and halo bias.
The halo model has seen success in the last few decades, and is used in modern analyses, albeit usually with some modifications, to model fully nonlinear scales (e.g. \cite{2019MNRAS.484..989W}).
However, the halo model struggles in the so-called ``transition regime'' between the one-halo and two-halo terms.
The halo model also fails to ensure large-scale conservation laws are satisfied, and as a result cannot be completely correct in its original form \cite{2016PhRvD..93f3512S}.
Despite the successes of both PT and the halo model in complementary regimes, it is clear that neither of these analytic models alone are adequate to fully describe the nonlinear density field.

Without sufficiently accurate analytic models of matter clustering, simulations can instead act as a model of nonlinear dynamics.
N-body simulations provide Monte Carlo realizations of Newtonian dynamics in the fully-nonlinear regime, producing a nonlinear matter density field for a given cosmological model.
As with PT and the halo model, N-body simulations are limited by a fundamental assumption - namely that evolution of the matter distribution is fully described by collisionless cold dark matter obeying Newtonian gravity.
However, this assumption does not limit the scales accessible to the model or the types of nonlinear structures it can produce, which are constrained only by numerical resolution.
Recent advances in computing have led to the rise of large-volume, high-resolution simulations (e.g. \cite{2018ApJS..236...43G, 2017ComAC...4....2P, 2009ApJ...705..156H, 2008ApJ...688..709T, 2019ApJ...875...69D}), albeit with questions of convergence at the percent-level \cite{2016JCAP...04..047S}. 
With these have come approximate methods of simulation that aim to obtain comparable solutions with much less computation time \cite{2016MNRAS.463.2273F, 2013JCAP...06..036T,2014MNRAS.437.2594W}.
As redshift surveys push to larger volumes and higher number density, the intractability of running many sufficiently resolved simulations at multi-Gpc volumes (necessary for capturing large modes and estimating covariance) to produce two-point statistics has motivated various fitting functions and interpolations of two-point statistics produced by high resolution simulations (e.g. \cite{2019MNRAS.484..989W,2019ApJ...874...95Z,2017ApJ...847...50L,2019ApJ...884...29N,2012ApJ...761..152T}).
At the smallest scales probed by observations, baryonic effects on the matter distribution are also a concern \cite{2004APh....22..211W,2004ApJ...616L..75Z}, and to properly simulate their impact on large-scale structure requires a full understanding of galaxy formation and hydrodynamic simulations that include feedback \cite{2019OJAp....2E...4C,2020MNRAS.491.2424V}.  
There have been some recent efforts to correct for baryons and mitigate this issue by modifying the output of dark-matter-only simulations \cite{2015MNRAS.454.1958M,2015JCAP...12..049S,2019JCAP...03..020S}.

In practice we do not observe the nonlinear matter density field, but instead its tracers.
Modeling the connection between tracers and the underlying density field is a complex task, and there are several prevailing approaches to this problem.
The large-scale bias approach extends the philosophy of perturbation theory to parameterize the tracer field as a linear combination of locally-leading gravitational observables \cite{2018PhR...733....1D}. 
These models have been successful on large scales, but face the same issue of PT for modeling dark matter, namely the presence of a nonlinear scale that characterizes the dark matter dynamics \cite{2015PhRvD..91b3508V}.
On top of this fundamental limit, there is an additional limiting scale, the nonlocality scale, which characterizes the formation of the tracer (e.g. halos/galaxies) and is not necessarily coincident with the nonlinear scale.
Recently, some pragmatic semi-analytic models have partially circumvented this issue by considering bias with respect to a fitting function for the non-linear matter power \cite{2020PhRvD.102l3522P}, using bias templates measured from simulations \cite{2020MNRAS.492.5754M}, or by extending the halo model via functions fit to simulations \cite{2020arXiv201108858M}.
In the halo occupation model approach, luminous tracers are assigned to a catalog of dark matter halos via a prescription for stochastically populating the halos - this is the Halo Occupation Distribution (HOD) framework \cite{2002ApJ...575..587B,2000MNRAS.318..203S,2017ApJ...848...76Z}. 
HOD models, usually combined with halos found in N-body simulations, are used for some modern analyses that include small-scale galaxy clustering \cite{2019MNRAS.484..989W,2014MNRAS.444..476R}.
There are also more involved models of the galaxy-halo connection that are less frequently used in cosmological analyses \cite{2018ARA&A..56..435W}.
Finally, hydrodynamic simulations that include baryonic/gas physics attempt to model galaxy formation more directly, albeit with stochastic subgrid models mixed in \cite{2019OJAp....2E...4C}.
These simulations are extremely computationally expensive to run, and typically cannot be run at the sufficient number of realizations or volume to be relevant for modern redshift surveys.

Given the limitations of existing models for the nonlinear density field and its tracers, a lofty goal is then to produce an interpretable, analytic model that is accurate across all scales of cosmological interest.
Toward this goal we build upon the hybrid PT-halo model approach put forward by \cite{2014MNRAS.445.3382M} (MS14) and \cite{2015PhRvD..91l3516S} (SV15) for modeling dark matter two-point correlators - Halo-Zeldovich Perturbation Theory (HZPT). 
There have been several other efforts in this hybrid-modeling direction for dark matter only correlators \cite{2011A&A...527A..87V,2016PhRvD..93f3512S,2020PhRvD.101l3520P,2013PhRvD..88h3524V}.
However, both HZPT in its original incarnation and these previous works do not account for baryonic effects on the matter two-point correlators or the more observationally-relevant modeling of general tracer two-point correlators.
We address these shortcomings in this paper.

The purpose of this paper is twofold. 
First, we aim to provide a fast, analytic, and accurate model for two-point correlators of matter (accounting for baryonic effects) and tracers on small scales, which necessitates modeling of halo exclusion and satellite galaxies. 
In addition, we provide power-law fits for matter correlators, and a joint distribution of cosmology and HZPT parameters for describing LRG-type (mock) galaxies as the foundation for an emulator-like approach to analysis of two-point statistics. 
We restrict our attention to real space correlators, as the most immediate application of HZPT is to projected statistics in a "3x2pt"-style analysis.

We first describe the N-body simulations and HOD mocks used in this paper in Section \ref{section:N-body}.
We review the HZPT model and outline its basic structures in Section \ref{section:review_hzpt}.
We apply the model to dark matter correlators and discuss the impact of baryons in Section \ref{section:matter_baryon}.
We apply the model to halos in Section \ref{section:halos}, and provide a detailed discussion of halo exclusion before moving on to mock galaxies in Section \ref{section:galaxies} and concluding in Section \ref{section:conclusions}.

\section{N-body Simulations and Halo Occupation Distribution}
\label{section:N-body}

\subsection{CrowCanyon simulations and CM HOD mocks}
We use particle output and halo catalogs from the \texttt{CrowCanyon} N-body simulations to test the HZPT model on matter and halo statistics. 
These simulations were run using FastPM \cite{2016MNRAS.463.2273F} with $N_{p} = 6144^{3}$ particles, a box size of $L_{\rm{box}} = 3200 \ \rm{Mpc}/h$, a boost factor $B=2$ (for a $12288^{3}$ PM force grid and Nyquist wavenumber $k_{\rm{Nyq}} \approx 12~\iM$), using the Planck15 cosmological parameters \cite{2016A&A...594A..13P} (without neutrino effects). 
\texttt{CrowCanyon} halos were identified using the \texttt{nbodykit} FoF halo finder \cite{1985ApJ...292..371D} with linking length $b=0.2$.
We computed simulation power spectra using \texttt{nbodykit} \cite{2018AJ....156..160H} with a FFT mesh using $N_{\rm{mesh}} = 2048$ ($k_{\rm{Nyq}} \approx 2 ~h/\rm{Mpc}$) using a correction for compensation \cite{2005ApJ...620..559J}, and a Triangular Shaped Cloud interpolation window.
Similarly, we computed correlation functions using FFTs on large scales with $N_{\rm{mesh}} = 1024$, matched at $r=10 \ \rm{Mpc}/h$ to the result of the \texttt{corrfunc} \cite{2020MNRAS.491.3022S} pair counting algorithm (as included in \texttt{nbodykit}) with 100 logarithmically-spaced bins and a maximum bin of $10\ \rm{Mpc} /h$.
Power spectra are sample variance cancelled using unitary amplitude (``paired-fixed'') power spectra \cite{2018ApJ...867..137V} at the same random seed as the N-body simulation initial conditions. 
To compute the linear theory power spectrum we use \texttt{CLASS} \cite{2011JCAP...07..034B}.
To quickly compute the Zeldovich power spectrum we modify a version of the FFTLog-based code employed in \cite{2016JCAP...12..007V} and \cite{2020arXiv200500523C}.
Fits are performed using the \texttt{scipy} implementation of the “Trust Region Reflective” optimization algorithm.

\begin{figure}[h!]
\center
\includegraphics[width=4. in, angle=0]{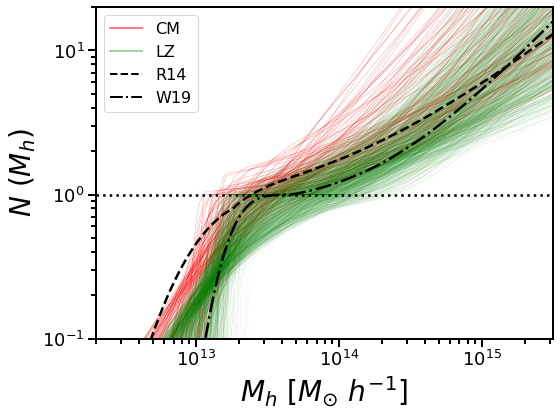}
\caption{Halo occupation space. CM refers to the wider HOD sample space corresponding to the larger-volume \texttt{CrowCanyon} simulations, while LZ refers to the LOWZ-like HOD sample space corresponding to the Aemulus simulations.
R14 and W19 refer to the halo occupations using the (fiducial) mean constrained HOD parameters (using the model of \cite{2007ApJ...667..760Z}) for the BOSS CMASS and LOWZ data from \cite{2014MNRAS.444..476R} and \cite{2019MNRAS.484..989W}, respectively.
\label{fig:nbody_hod_space}}
\end{figure}

We used the \texttt{nbodykit} \cite{2018AJ....156..160H, 2017AJ....154..190H} implementation of the simple 5-parameter Zheng ’07 HOD model \cite{2007ApJ...667..760Z} to populate \texttt{CrowCanyon} halos with galaxies.
This implementation modulates the satellite occupation by that of the centrals, assumes no halo-central mis-centering, and places satellite galaxies in halos according to an NFW profile.
We use 100 sets of HOD parameters sampled from a symmetric latin hypercube with a number density fixed to near the BOSS CMASS \cite{2016MNRAS.455.1553R} value ($\bar{n}_{g} = 4.2 \times 10^{-4} \ h^{3} \ \rm{Mpc}^{-3}$) at $z=0.55$ (Fig.~\ref{fig:nbody_hod_space}). 
The parameter $\log M_{\rm{min}}$ is not drawn from the hypercube, and is instead fixed by integrating over the \texttt{CrowCanyon} halo mass function to reproduce the appropriate $\bar{n}_{g}$.
The ranges of parameters considered are: $\alpha \in \left[0.5 , 1.0 \right]$, 
$\log M_{1} \in  \left[13.5 , 14.5\right]$, $\log M_{0} \in  \left[11 , 13.5\right]$, $\sigma_{\log M} \in  \left[ 0.01, 0.8 \right]$.
This results in a large range of satellite fractions ($f_{\rm{sat}} \approx 0.01-0.6$), which is discussed further in Section \ref{section:galaxies}.
We refer to this HOD mock galaxy sample as the ``CM'' sample since it approximates the BOSS CMASS galaxy number density, redshift, and roughly follows the HOD parameterization of the CMASS analysis of \cite{2014MNRAS.444..476R}.

\subsection{Aemulus simulations and LZ HOD mocks}
We similarly generate LOWZ-like HOD mock catalogs from the Aemulus simulations \cite{2019ApJ...875...69D}.
We use 10 different cosmologies (Boxes 0-9) with 20 HODs per cosmology.
We use a number density close to the LOWZ value ($\bar{n}_{g} = 3 \times 10^{-4} \ h^{3} \ \rm{Mpc}^{-3}$), and use a snapshot at $z=0.25$.
Halos are populated according to the 200b mass definition of the NFW radius and concentration.
We computed simulation power spectra using \texttt{nbodykit} \cite{2018AJ....156..160H} with a FFT mesh using $N_{\rm{mesh}} = 1024$ using a correction for compensation \cite{2005ApJ...620..559J}, a Triangular Shaped Cloud window, and interlaced anti-aliasing \cite{2016MNRAS.460.3624S}.
We extend the maximum \texttt{corrfunc} pair count bin to $20 \ \M$ and use an \texttt{FFTCorr} grid of $512^{\mathbf{3}}$ for correlation function measurements.
We use a similar range of parameters as \cite{2019MNRAS.484..989W}: $\alpha \in \left[0.5 , 1.5 \right]$, $\frac{M_{1}}{M_{\rm{min}}} \in  \left[2.5 , 20.0 \right]$, $\frac{M_{0}}{M_{1}} \in  \left[0.0 , 0.4\right]$, $\sigma_{\log M} \in  \left[ 0.01, 0.8 \right]$.
For each set of HOD parameters, the minimum halo mass parameter $\log M_{\rm{min}}$ is fixed to match the LOWZ number density once all of the other parameters have been chosen.
This value is determined by an integral over the halo mass function - to account for variation of the mass function with cosmology, we use the Aemulus emulator for the halo mass function trained on the simulations \cite{2019ApJ...872...53M}.
We refer to this HOD mock galaxy sample as the ``LZ'' sample since it approximates the BOSS LOWZ galaxy number density, redshift, and HOD parameterization of the LOWZ analysis of \cite{2019MNRAS.484..989W}.
Figure \ref{fig:nbody_hod_space} shows the halo occupations corresponding to these parameters (as well as the same for the CM sample).

\section{Review of Halo-Zeldovich Perturbation Theory}
\label{section:review_hzpt}

In this section, we review the HZPT model for two-point statistics as developed in MS14, SV15, 
and \cite{2017JCAP...10..009H} (H17), briefly recounting the relevant aspects of the halo model. 
We discuss each of the terms, their PT/halo model origin, and the scales at which they are relevant.
Previous iterations of the HZPT model were applied only to matter and halo-matter correlators, and discussion in this section is restricted to these models, though we expand upon them more generally for tracers in Sections \ref{section:halos} and \ref{section:galaxies}.

\subsection{The halo model}
\label{sec:halo_mod}
The halo model \cite{2000MNRAS.318..203S,2000MNRAS.318.1144P,2002PhR...372....1C,2000ApJ...543..503M} makes the assumption that all matter resides in virialized halos of mass $M_{\rm{vir}} = \frac{4}{3} \pi R_{\rm{vir}}^{3} \Delta_{\rm{vir}} \bar{\rho}_{m}$, and splits the two-point statistics of the matter field into correlations between halos (two-halo term) and correlations within a single halo (one-halo term):
\begin{equation}
P_{mm}(k) = P_{1h}(k) + P_{2h}(k),\ \xi_{mm}(r) = \xi_{1h}(r) + \xi_{2h}(r).
\end{equation}
The ingredients of the halo model are: the halo mass function $dn(M)$ with $n(M)$ the number density of halos at fixed mass $M$, along with halo bias $b(M)$, and the spherically averaged halo profile $\rho_{M} (r)$(e.g. NFW \cite{1996ApJ...462..563N}).
The usual halo-model one-halo and two-halo expressions are then:
\begin{equation}
P_{1h}(k) = \int dn(M) \frac{M}{\bar{\rho}} |u_{M}(k)|^{2}, 
\label{eqn:1halo}
\end{equation}
\begin{equation} 
P_{2h}(k) =  \left( \int dn(M) b(M) u_{M}(k) \right)^{2} P_{L}(k),
\label{eqn:2halo}
\end{equation}
with linear power $P_{L}(k)$, and where the configuration space quantities are given by the Fourier transform of eqns. \ref{eqn:1halo} and \ref{eqn:2halo}.
Here $u_{M}(k)$ is the Fourier transform of the density profile normalized by the mass enclosed in the halo:
\begin{equation}
u_{M}(k) = \frac{4 \pi}{M} \int_{0}^{R_{\rm{vir}}} dr \ r^{2} \rho_{M}(r) j_{0}(kr),
\label{eqn:kprofile}
\end{equation}
with $j_{0}$ denoting the $0$th-order spherical Bessel function.
The profile is usually parameterized in terms of a characteristic scale radius $r_{s}$, (which can be written in terms of the halo concentration $c_{\rm{def}}(M)=\frac{R_{\rm{def}}}{r_{s}}$):
\begin{equation}
    \rho_{M}(r) = \frac{\rho_{0}}{\left(1+\frac{r}{r_{s}}\right)^{2}\left(\frac{r}{r_{s}}\right)}.
    \label{eqn:halo_profile}
\end{equation}
We only explicitly compute halo model quantities in Appendix~\ref{appendix:baryonification}, and in that case use the NFW concentration-mass relation of Ref.~\cite{2014MNRAS.441.3359D} and the 200c mass definition.

\subsection{Two-halo - Zel'dovich}
\label{sec:two_halo}
The HZPT model replaces the traditional halo model two-halo term (eqn.~\ref{eqn:2halo}) with the Zel'dovich Approximation (ZA), or the leading-order Lagrangian perturbation theory (LPT) power spectrum\footnote{Using an alternative perturbative substitute for the two-halo term can also prove fruitful, and we provide an example in Section \ref{sec:alt_two_halo}, but by default in this paper we stick with the ZA.}  \cite{1970A&A.....5...84Z,2014MNRAS.439.3630W,2008PhRvD..77f3530M,2014JCAP...06..012T}. 
The ZA provides a beyond-linear-theory description of large scales, including large-scale nonlinear bulk flows.
A benefit of the ZA is that the IR resummation that would be necessary in Eulerian perturbation theory (SPT) or in an EFT extension thereof is not required at several-percent accuracy, \cite{2020PhRvD.101l3520P, 2015PhRvD..91b3508V, 2012JCAP...07..051B}, as the Baryonic Acoustic Oscillation (BAO) wiggles are already captured quite well by the ZA \cite{2016JCAP...03..057V}.
More importantly for the purpose of this paper, 
ZA provides a useful ansatz for halo model 
extension, since a compensated halo profile added to 
ZA provides a description which is consistent 
with perturbation theory in the regime of 
its validity on large scales (Section II of SV15). 
To address the known deficit in power in the ZA power spectrum on large scales, SV15 matched the ZA power spectrum to that of SPT to write the amplitude of the one-halo term that also describes small scales, which we now address.

\subsection{One-halo - Broadband Beyond Zel'dovich}
\label{sec:one_halo}
The central feature of the small-scale HZPT model is the Broadband Beyond Zel'dovich (BB) term. 
In its initial formulation (MS14), this term replaces the one-halo term (eqn~\ref{eqn:1halo}) to express a contribution to the power that is provided by an expansion of the Fourier transform of the halo profile in even powers of comoving wavenumber $k$.
The coefficients of this expansion can in principle be obtained by integrating $r^{2n}$-moments of a prescribed halo profile up to a chosen “halo radius” at which the profile is truncated, though this was not done in MS14 or SV15 and these coefficients were simply fitted to simulations (expressions given in Appendix \ref{appendix:theory}).
To prevent large-$k$ divergences, SV15 took this expansion in even powers of $k$ and replaced it with a Pad\'e-type term:
\begin{equation}
P_{BB}(k) = A_{0} F(k) \frac{\sum_{m=0}^{m=n_{\rm{max}}-1} (k R_{m})^{2m}}{\sum_{n=0}^{n=n_{\rm{max}}} (k R_{nh})^{2n}}
\label{eqn:pade}
\end{equation}
The Pad\'e approximation to the the $k^{2}$ expansion improves the range of validity of the model greatly by forcing the expansion to smoothly transition to zero as $k \rightarrow \infty$.
Not only does this resummation remove high-$k$ divergence due to polynomial terms, but it also increases the maximum wavenumber $k$ up to which the BB term is a decent approximation to the Fourier transform of an idealized halo profile, which also transitions to zero for large $k$. 
We note that in the EFT sense, the BB term is not “stochastic”, since we do not enforce that it is uncorrelated with the ZA term.
The two-point correlation function (2PCF) is given by the Fourier transform of this expansion, which is analytic\footnote{To clarify, we take ``analytic'' in the sense of closed-form (as in \cite{2000MNRAS.318..203S}), rather than in the technical sense of functions.}, and an expression for which (for $n_{\rm{max}}=0,1,2$) can be found in Appendix \ref{appendix:theory}.

\subsection{Compensation}
\label{sec:compensation}
A well-known limitation of the original halo model is that mass and momentum are not conserved on large scales \cite{2016PhRvD..93f3512S}. 
In the $k\to0$ limit, conservation of these quantities requires $\lim_{k \to 0} P_{mm} \propto k^{4}$ \cite{1980lssu.book.....P}.
The violation of this requirement for the halo model arises from the $k^{0}$ contribution of the one-halo term, which is due to the Poisson contribution from a finite number of halos \cite{2000MNRAS.318..203S}.
The HZPT model addresses this by multiplying the BB Pad\'e expansion term by a compensation kernel $F(k)$, which suppresses the low-$k$ $k^{0}$ contribution (though at leading order this term goes like $k^{2}$).
It is possible to explore more complicated forms of the compensation kernel (MS14), or to compensate the functional form of the halo profile iteslf \cite{2020PhRvD.101j3522C}, but here we keep with the previous iterations of HZPT and use the simple Lorentzian kernel with a single parameter $R$:
\begin{equation}
\label{eqn:compensation}
F(k) = \left( 1 - \frac{1}{1 + k^{2} R^{2}} \right).    
\end{equation}
SV15 matched to SPT to find a value of the compensation parameter $R$ that was in good agreement with simulation measurements ($R \approx 26 \ \mathbf{[Mpc/h]}$ at $z=0$).
We discuss compensation for tracer-matter cross-correlations in Appendix \ref{appendix:halos_comp}.

\subsection{The full model}
\label{sec:full_model}
\begin{figure}[h!]
\center
\includegraphics[width=2.95 in, angle=0]{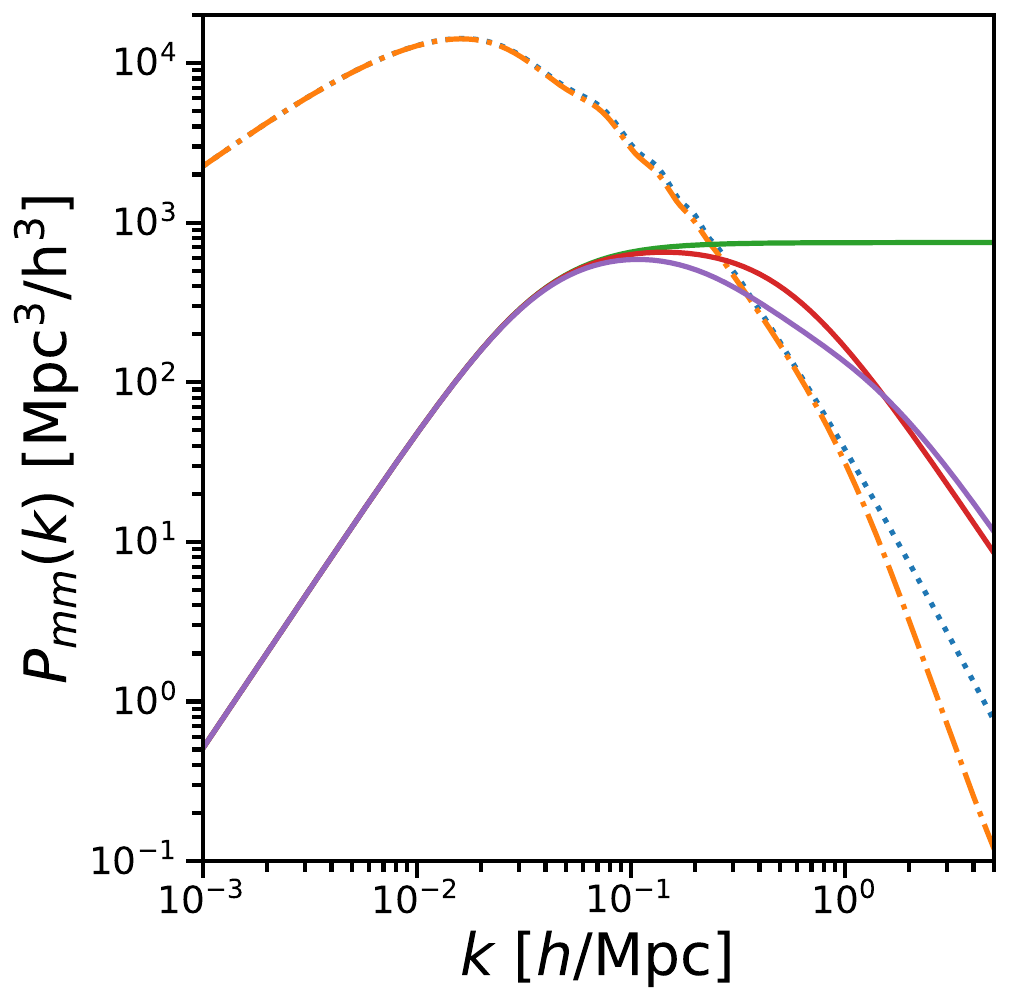}
\includegraphics[width=2.95 in, angle=0]{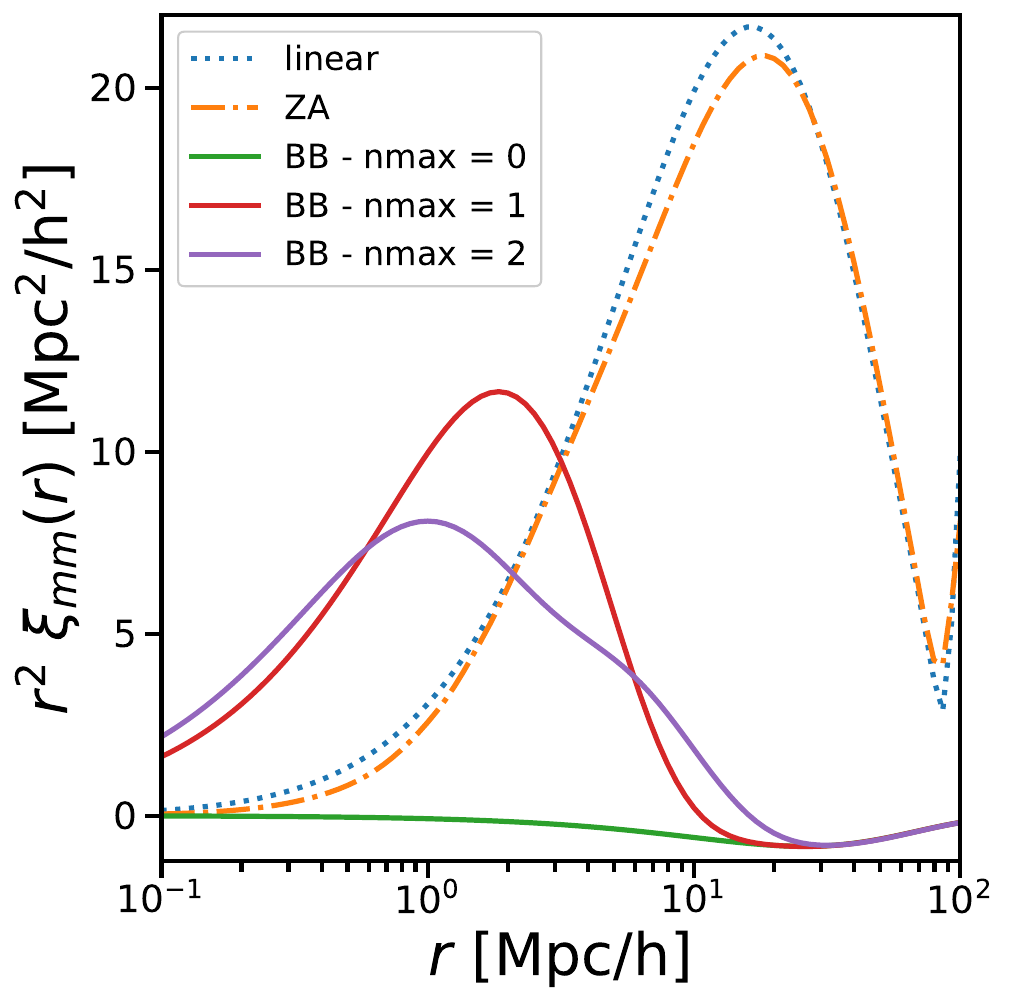}
\caption{\textit{Left:} Illustration of separate components of our model in Fourier space, as well as a comparison to linear theory.
\textit{Right:} The same components of the model in configuration space.
\label{fig:hzpt_model}}
\end{figure}

The contributions to the model from the ZA and different BB terms are shown in Figure \ref{fig:hzpt_model}.
The parameters are: $A_{0}$, which is related to the one-halo amplitude $\bar{\rho}_{m}^{-2} \int dn(M) M^{2}$ and does not depend on the profile, the $R_{nh}$, which are associated to the $r^{2n}$-moments of the halo profile, and the compensation scale $R$.  
A detailed discussion of these terms is provided in Appendix \ref{appendix:theory}. 
There we also provide a full review of the original machinery of MS14 for the expansion in even powers of $k$.
In the main text we will take a pragmatic approach, always fitting for the model parameters.

\section{Matter Correlators \& Baryons}
\label{section:matter_baryon}
We present the HZPT model for real-space two-point correlators of matter, including the effects of baryonic physics, which are most relevant for weak lensing observations.
We review a calculation of MS14 of the profile expansion coefficients in the presence of a single model of AGN feedback, before performing an expanded calculation in the context of HZPT using several baryonic physics models. 
We also present a higher-$n_{\rm{max}}$ model that fits down to $k \approx 8 \ h/\rm{Mpc}$ at the 3\% level, as well as a model with an augmented two-halo term that fits close down to $k \approx 10 \ h/\rm{Mpc}$ at the 1\% level.

\begin{figure}[h!]
\center
\includegraphics[width=2.95 in, angle=0]{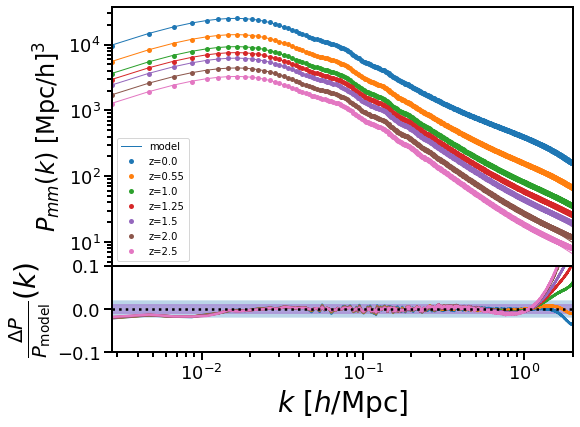}
\includegraphics[width=3.05 in, angle=0]{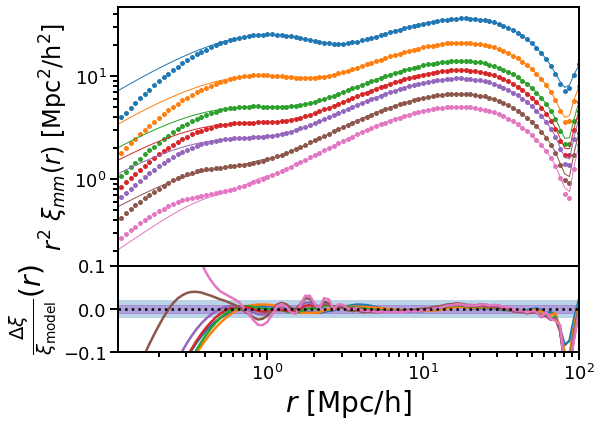}
\caption{\textit{Left:} Fits to the \texttt{CrowCanyon} matter power spectrum using the base HZPT model for a range of redshifts ($z=0-2.5$).
\textit{Right:} The same for the matter correlation function.
Colored bands are at $1\%$ and $2\%$.
\label{fig:matter_z}}
\end{figure}

Before turning to baryons we briefly remark on the use of HZPT as a model of dark matter two-point correlators.
Slightly generalizing the results of SV15, we show the fits over a range of redshifts ($0\leq z \leq 2.5$) of the HZPT model to the power spectrum and correlation function in the \texttt{CrowCanyon} simulations in Figure \ref{fig:matter_z}.
The fits for $P_{mm}(k) ~(\xi_{mm}(r))$ are accurate to $\approx 2\%$ for $k>1~\iM$ $(r<1~\M)$ for all but the highest redshift considered.
Fits are performed in Fourier space, and these best-fit parameters are provided as input to the Fourier space model.

\subsection{Modeling baryonic physics}
\label{section:baryons}
To avoid biasing inferred cosmological parameters obtained through the matter two-point correlators, we must account for the effect of baryons on the matter distribution. 
Baryons modify the dark-matter-only (DMO) halo profile in several ways, which has been explored in detail (e.g. \cite{2016JCAP...04..047S,2019OJAp....2E...4C,2018MNRAS.480.3962C,2019JCAP...03..020S, 2020MNRAS.491.2424V}). 
In the presence of baryons, gas and stars must be accounted for in the halo profile in addition to dark matter.
AGN and supernova feedback effects also move gas away from the halo center, which redistributes dark matter within the halo profile.
From the perspective of the HZPT model, the effect of baryons should only be to modify the BB terms (see Appendix \ref{appendix:baryonification} for an illustrative calculation using halo profiles). 
Baryons should not affect the large scales relevant for the halo compensation $R$ or the one-halo amplitude $A_{0}$, assuming conservation of mass between the DMO and DM+baryon scenarios (up to two-fluid corrections in the ZA \cite{2019JCAP...06..006C,2016PhRvD..94f3508S,2021MNRAS.503..406R,2020JCAP...02..005B}).
This is the same rationale used to motivate scale cuts (e.g. \cite{2018PhRvD..98d3528T}).
MS14 fitted changes in the power spectrum due to a single AGN model of feedback in the coefficients for the first three terms of the $k^{2}$ profile expansion up to $k=0.8\ h/\rm{Mpc}$.
They found that the change in $A_{0}$ (which is the same as our $A_{0}$ up to small changes due to the compensation term) is almost an order of magnitude lower than changes in the higher-order parameters, which change at the $\sim 5\% $ level, and if they fix $A_{0}$ the change in the other terms is larger but still effectively captured by the profile expansion. 

\begin{figure}[h!]
\center
\includegraphics[width=4 in, angle=0]{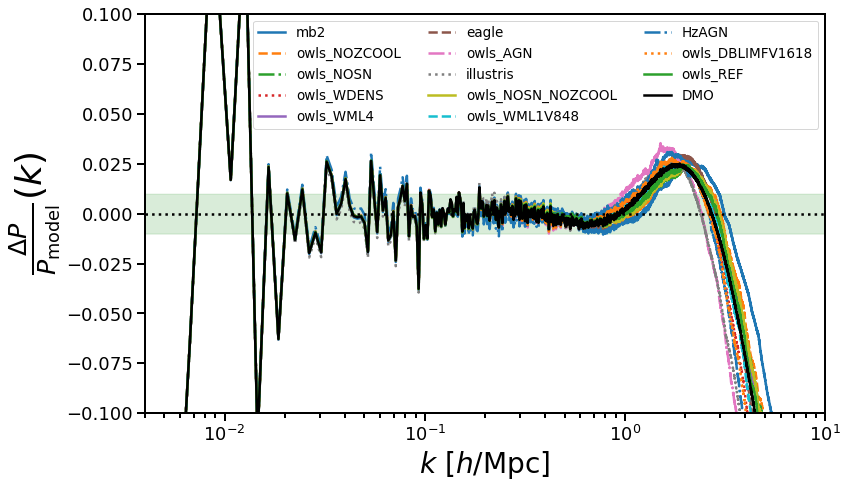}
\includegraphics[width=3.85 in, angle=0]{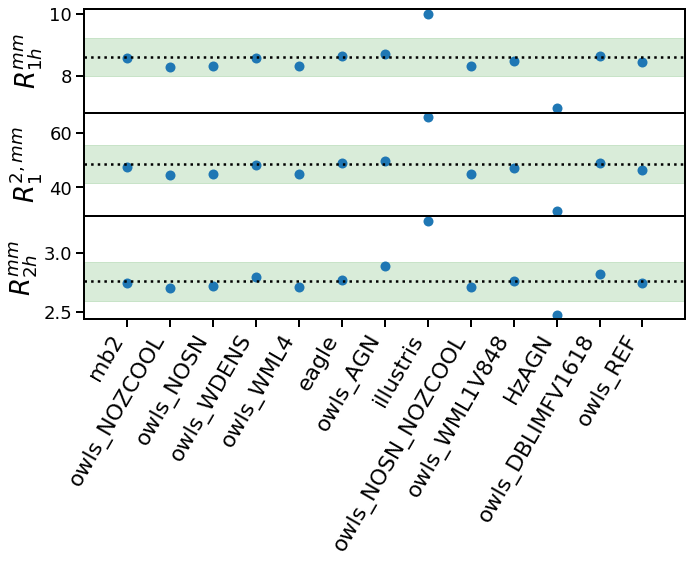}
\caption{\textit{Top:} Residuals for the power spectrum at $z=0$ for all baryonic models considered. 
Shaded area corresponds to a 1\% deviation.
\textit{Bottom:} HZPT parameter values fit for the different feedback models. DMO fits are denoted by the dotted lines and shaded areas denote rms deviations from the DMO case.
\label{fig:matter_baryon_resid}}
\end{figure}

We achieve a similar but improved result compared to MS14 using a more involved comparison.
We use a larger $k$-range, fitting out to $k=1\ h/\rm{Mpc}$ using the matter power spectrum from simulations.
We also employ a more diverse range of baryonic physics models by multiplying the dark-matter only \texttt{CrowCanyon} matter power at $z=0$ by the ratio $P_{\rm{baryon}}/P_{\rm{DMO}}$ for 13 different models from \cite{2011MNRAS.415.3649V,2014MNRAS.444.1518V,2014MNRAS.444.1453D,2015MNRAS.450.1349K,2010MNRAS.402.1536S,2015MNRAS.446..521S,2018MNRAS.480.3962C} as used in \cite{2019MNRAS.488.1652H}.
Figure \ref{fig:matter_baryon_resid} shows fits to dark-matter-only power spectra and power spectra including the effects of baryons for several models with $A_{0}$ and $R$ fixed to their DMO values. 
Clearly the $R_{n,nh}$ parameters are flexible enough to accurately account for feedback, and marginalizing over them should remove biases in cosmological parameter constraints.
We note that while all models are fit to the $1\%$ level down to $k=1\ h/\rm{Mpc}$, the simulations with the largest deviations in the $R_{nh}$ parameters from the DMO case are Illustris and Horizon-AGN.
This is the case for Illustris since it has been shown to have an unrealistically strong feedback model (compared to other hydrodynamic simulations) in terms of its effects on the power spectrum for $k\leq 1$ due to low baryon fraction (compared with the observed value in galaxy groups) in high mass halos \cite{2020MNRAS.491.2424V}.
For Horizon-AGN, there is a large-scale $1\%$ excess of the $P_{\rm{baryon}}/P_{\rm{DMO}}$ ratio above unity that causes a relatively large change in the HZPT $R_{nh}$ parameters (the source of this deviation is discussed in Appendix A of \cite{2018MNRAS.480.3962C}, and may not be physical).

\subsection{Extended power spectrum model}
\label{section:ext_pmm}
We briefly explore two extended models, which probe smaller scales where the effects of baryons are stronger, as modern cosmic shear measurements probe angular scales that receive contributions from these length scales.
The first extension focuses on the one-halo term, as well as cosmology dependence of those parameters, and the second focuses on the two-halo term.
We compute the power spectrum using \texttt{nbodykit} as described in Section \ref{section:N-body} but with a finer mesh in the call to \texttt{FFTPower} with $N_{\rm{mesh}}=10240$, and additional interlaced anti-aliasing \cite{2016MNRAS.460.3624S}.
This grid corresponds to a $k_{\rm{Nyq}} \approx 10 \ h/\rm{Mpc}$ and using the above settings should be trustworthy out to this scale \cite{2018AJ....156..160H}.

\subsubsection{Extending the one-halo term}
\begin{figure}[h!]
\center
\includegraphics[width=4. in, angle=0]{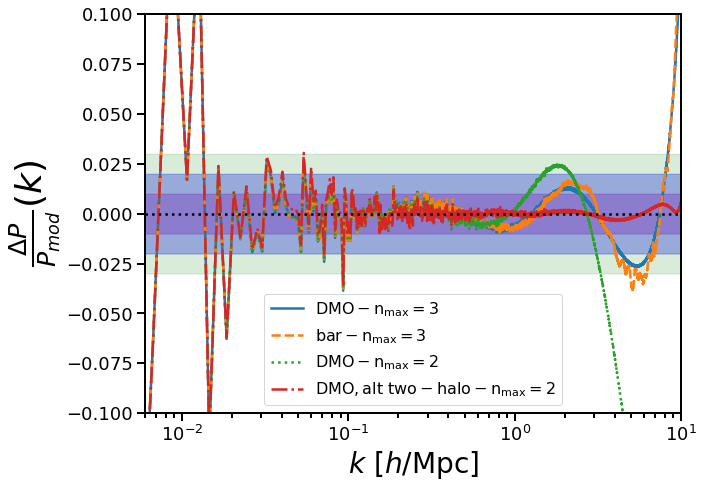}
\caption{Fits to the matter power spectrum for dark-matter only in our simulations as well as with baryonic feedback effects included through the power spectrum ratio $P_{\rm{bar}}/P_{\rm{DMO}}$ for Illustris, the most extreme feedback model we consider, at $z=0$ using the high-k extended model with $n_{\rm{max}} = 3$.
We also show the alternate two-halo HZPT model described in Section \ref{sec:alt_two_halo} (red line).
Colored regions are shown at 1,2, and 3\%.
\label{fig:matter_power_z_illustris}}
\end{figure}

We extend the model for $P_{mm} (k)$ to include one higher-order BB term ($n_{\rm{max}}=3$) to get to $3\%$-level accuracy out to $k \approx 8 \ h/\rm{Mpc}$.
We see an upturn in Figure \ref{fig:matter_power_z_illustris} beginning at $k=8 \ h/\rm{Mpc}$ which the model fails to fit.
We see that the $R_{nh}$ parameters can account for the strongest baryonic feedback (Illustris), which is perhaps not surprising given the fact that we have added two parameters - which are interpretable as the $k^{6}$ expansion coefficient reprocessed through the Pad\'e expression.

\begin{figure}[h!]
\center
\includegraphics[width=3. in, angle=0]{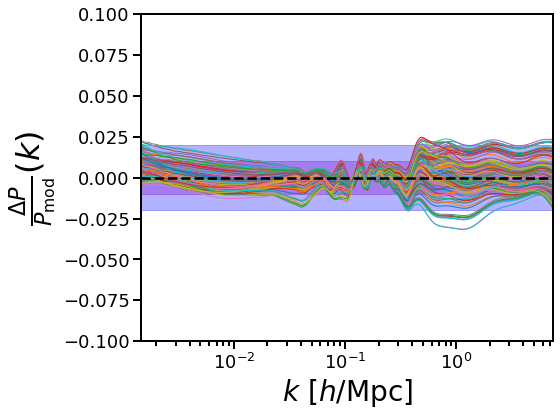}
\includegraphics[width=3. in, angle=0]{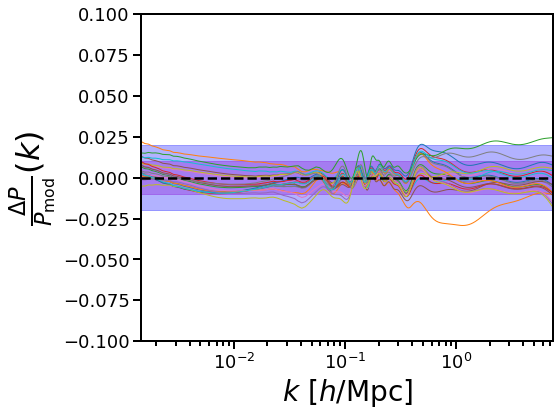}
\caption{\textit{Left:} Residuals of power law fits in the training set of 100 values of $\sigma_{8}$ and $\Omega^{\rm{EMU}}_{m} = \Omega_{cb}$.
\textit{Right:} Residuals of power law fits in the test set of 20 values of $\sigma_{8}$ and $\Omega^{\rm{EMU}}_{m}$.
Colored bands show 1 and 2 \%.
See Fig.~\ref{fig:hmcode} for a comparison to \texttt{HMCode2020}.
\label{fig:matter_nmax3_cosmo_test_train}}
\end{figure}

Following SV15, we here provide fitted power-law dependence of all $n_{\rm{max}}=3$ parameters for the $P_{mm}(k)$ (DMO) model described in the previous section at $z=0$ (Eqn.~\ref{eqn:powlaws}). 
To quickly obtain matter power spectra for cosmologies with different values of the matter density parameter $\Omega_{m}$ and the matter density variance in spheres of $8 \ \rm{Mpc}/h$ $\sigma_{8}$, we use \texttt{CosmicEmu} \cite{2017ApJ...847...50L}, using 100 randomly generated power spectra with values of $\Omega_{cb} = \Omega^{\rm{EMU}}_{m} \in [.26,.34]$ and $\sigma_{8} \in [.7,.9] $ and 20 test spectra in the same range (Fig.~\ref{fig:matter_nmax3_cosmo_test_train}).
For the other emulator parameters we fix $h=0.6774$, $\Omega_{b}=0.0486$, $\Omega_{\nu}=0.0014$, $n_{s}=0.9667$ $w_{0}=-1$, and $w_{a}=0$ 

\begin{equation}
\label{eqn:powlaws}
    A_{0} = 777 \left(\frac{\sigma_{8}}{0.8}\right)^{4.33} \left(\frac{\Omega_{cb}}{0.3}\right)^{-1.83} 
\end{equation}
\begin{alignat*}{2}
R &= 25.3 \left(\frac{\Omega_{cb}}{0.3}\right)^{-0.58} \qquad &&R_{1h}= 8.56 \left(\frac{\sigma_{8}}{0.8}\right)^{2.34} \left(\frac{\Omega_{cb}}{0.3}\right)^{-2.19}  \\
R_{1} &= 7.34 \left(\frac{\sigma_{8}}{0.8}\right)^{2.37} \left(\frac{\Omega_{cb}}{0.3}\right)^{-1.39} \qquad &&R_{2h} = 2.93 \left(\frac{\sigma_{8}}{0.8}\right)^{1.56} \left(\frac{\Omega_{cb}}{0.3}\right)^{-1.24}   \\
R_{2} &= 1.99 \left(\frac{\sigma_{8}}{0.8}\right)^{1.16} \left(\frac{\Omega_{cb}}{0.3}\right)^{-0.96} \qquad
&&R_{3h} = 1.51 \left(\frac{\sigma_{8}}{0.8}\right)^{1.12} \left(\frac{\Omega_{cb}}{0.3}\right)^{-0.96} 
\end{alignat*}

The 3-parameter power law is accurate 
with rms residuals of 1\% or less for the test set on all scales.
In the context of this computation, this level of accuracy is competitive with state-of-the-art non-linear matter power spectrum models \cite{2021MNRAS.502.1401M}
- we provide a comparison to the right panel of Fig.~\ref{fig:matter_nmax3_cosmo_test_train} in Appendix~\ref{appendix:hmcode}.
The few large-$k$ residuals that go slightly past $2\%$ correspond to the most extreme values of $\Omega^{\rm{EMU}}_{m}$ at the edge of our range.
We find positive exponents for all parameters with respect to $\sigma_{8}$ and generally negative ones for the parameters with respect to $\Omega_{m}^{\rm{EMU}}$.
We note that fitted value for the exponent on the compensation scale $R$ is close to zero ($<10^{-15})$, so the value of $R$ is essentially independent of the value of $\sigma_{8}$, and so we treat $R$ only as a function of $\Omega_{m}^{\rm{EMU}}$.

\subsubsection{Alternate two-halo term}
\label{sec:alt_two_halo}
As mentioned in Section \ref{sec:two_halo}, one might consider alternatives to ZA for the two-halo term in the HZPT model, and we briefly explore such an extension here\footnote{We are grateful to Zvonimir Vlah for suggesting the main idea of this section.}.
One such alternative is based on the power due to the linear correlation function shifted by the ZA displacement:
\begin{equation}
    \label{eqn:plin_ir}
    P_{\mathrm{alt}}(\mathbf{k})  = 4 \pi \int dq q^{2} \xi_{L}(q) \mathrm{e}^{-k_{i} k_{j} A_{ij}(q)}
\end{equation}
where $\xi_{L}$ is the linear correlation function and $A_{ij}(q)$ is the LPT displacement difference cumulant \cite{2015JCAP...09..014V,2017JCAP...08..009M,2014MNRAS.439.3630W,2016JCAP...03..057V, 2013MNRAS.429.1674C}.
A benefit of taking $P_{\mathrm{alt}}$ as the starting point for a new two-halo term is that it remains easy to compute while preserving many of the advantages of ZA, and also provides slightly more power than ZA on quasi-linear scales.
This modified version of ZA can be further augmented in a manner similar to the models of \cite{2012JCAP...12..011T,2016JCAP...03..057V} by adding a transfer function style polynomial term multiplying $P_{\mathrm{alt}}$, such that
\begin{equation}
    \label{eqn:alt_hzpt}
    P_{\mathrm{alt, HZPT}}(k) = \left(1 + \alpha k^{2} + \beta k^{4} \right)P_{\mathrm{alt}}(k) + P_{BB}(k)
\end{equation}
where $\alpha, \beta$ are free parameters.
This model keeps with the HZPT spirit of an analytic Fourier transform and provides a converged inverse Fourier transform due to the Gaussian term of the ZA-like term\footnote{We find that fits to $\xi^{mm}$ using $\xi_{\mathrm{alt}}+\xi_{BB}$ with $\nm=2$ are also accurate at the percent level.}.
For $\nm=2$, this model has the same number of parameters as the $\nm=3$ model described in Section \ref{section:ext_pmm}.
The success of this model in fitting the \texttt{CrowCanyon} matter power spectrum is demonstrated by the red line in Fig.~\ref{fig:matter_power_z_illustris}.
Clearly for the same number of parameters, $P_{\mathrm{alt, HZPT}}$ outperforms the $\nm=3$ model, achieving 1\% residuals up to $k = 10~\iM$.
We also find that fixing $\beta=0$ results in similar performance to the $\nm=3$ model, with one fewer parameter.
This illustrates the power of the form of the HZPT model and that it is possible to further improve beyond the models we present in this paper  within the HZPT framework by more carefully balancing the work sharing between 
the HZPT two-halo and one-halo terms.

We anticipate that using the form of eqn.~\ref{eqn:alt_hzpt} may also improve the accuracy of tracer HZPT models, and also results in exact expression for tracers using linear bias.
We find that since for tracers we mostly limit our attention to wavenumbers below $1-2~\iM$, using ZA alone suffices for our purposes.
It would be quite interesting to further explore HZPT models of tracers based on $P_{\rm{alt}}$ or other similar two-halo terms.

\section{Halos and Exclusion}
\label{section:halos}
In this section we present the HZPT model for halo-halo and halo-matter two-point statistics in configuration and Fourier space.
While these quantities are not directly observable, understanding them is key to accurately modeling non-perturbative effects in the transition regime - namely, the discreteness of the halo field, and halo exclusion.

\subsection{Small-scale halo clustering}

The HZPT model was successfully applied to halo-matter cross-correlation two-point statistics, but was not applied to halo clustering (auto-correlation) statistics.
H17 showed that fits for the power spectrum and correlation function to simulations were accurate to 2\% to $ k=1\ h/\rm{Mpc}\  (r \sim 4 \ \rm{Mpc}/h $) between $z=0-1$ for several halo mass bins between $10^{13}-10^{14} M_{\odot}$.
The model H17 used was the base HZPT model for dark matter with $n_{\rm{max}}=2$ and with an additional linear bias parameter $b_{1}$. 
The HZPT parameters were fixed as fitted power laws of the linear halo bias $b_{1}$ and $\sigma_{8}$.
Allowing all parameters (including $b_{1}$) to be free, we find that we can produce fits  for these halo-matter correlators that are slightly more accurate on large scales
corresponding to those of H17 (our bins 6-8) and also for a wider range of halo-mass bins (see Table~\ref{tab:halo_table}) as shown in the right panels of Figs.~\ref{fig:xih_resid} and \ref{fig:Ph_resid}.
With the $R_{nh}$ free we are able to fit down to smaller scales that H17 struggled with in modeling the correlation function. 
H17 attributed this failure of the configuration-space model to one-halo effects since the dependence on profiles cannot be fully described with a simple power law in $b_{1}$, and suggested a more complete treatment of nonlinear/nonlocal bias. 
We examine where the model fails and succeeds in more detail in the next section, but first turn to the non-perturbative features present in halo auto-correlations.

Non-perturbative modeling is necessary to accurately capture small-scale halo clustering.
Ref \cite{2013PhRvD..88h3507B} (B13) conducted a detailed study of halo exclusion and halo auto-correlation stochasticity using N-body simulations.
Manufacturing a discrete halo field from the continuous matter field introduces Poisson noise from the finite number of resulting objects in a given volume.
This contributes at zero-lag in configuration space and on all scales in Fourier space as the well-known ``Poisson shot noise’' or fiducial stochasticity.
In the $k \to 0$ limit, however, because the tracer field is discrete, there is a constant contribution to the power spectrum that involves an integral over the correlation function.
This means that the constant noise on all scales in Fourier space may be sub- or super-Poisson.
The sub/super-Poisson noise has been investigated in detail  \cite{2010PhRvD..82d3515H, 2019PhRvD.100d3514S, 2007PhRvD..75f3512S, 2013PhRvD..88h3507B}.

In addition to the scale-independent contribution from discretizing the field, the phenomenon of halo exclusion introduces a scale-dependent contribution to the two-point statistics.
Halo exclusion follows directly from the foundational assumption of the halo model - that all matter is contained in non-overlapping collapsed dark matter halos.
If halos are idealized as spherical, the phenomenon of exclusion appears quite straightforward.
Since halos may not overlap, it is not possible for the (spherical) halo field to be correlated on scales below the sum of halo radii.
This is reflected in a discontinuous drop to a value of -1 in the halo correlation function at the exclusion scale.
For this simplified case of spherical halos at fixed mass, we can write the following expressions of B13 for the discrete (auto) correlation function for halos:
\begin{equation}
 \xi_{hh}^{(d)} (r) - \frac{1}{\bar{n}}\delta^{(D)}(\mathbf{r}) =
 \begin{cases} 
      -1 & r < R_{\rm{exc}} \\
      \xi_{hh}^{(c)} (r) & r \geq R_{\rm{exc}},
 \end{cases}
\label{eqn:exclusion_xi}
\end{equation}
or, writing the two-point function, we have $ \xi_{hh}^{(d)} (r) - \frac{1}{\bar{n}}\delta^{(D)}(\mathbf{r}) = (\xi_{hh}^{(c)} (r) + 1) \Theta_{H} (r-R) -1 $, where $R_{\rm{exc}} = R_{\rm{exc}} (M)$ is the exclusion scale, and in the notation of B13, $(d)$ signifies “discrete” as to be distinguished from $(c)$ “continuous”\footnote{Here and in the remainder of this work, we take $\xi^{(c)}$ to contain any nonlinear or non-perturbative clustering outside the exclusion scale.}.

For the power spectrum the corresponding expression is:
\begin{equation}
P_{hh}^{(d)} (k) = \frac{1}{\bar{n}} + P_{hh}^{(c)} (k) - V_{\rm{excl}} \left( W_{R} (k) + \left[ W_{R} * P_{hh}^{(c)} \right] (k)  \right),
\label{eqn:exclusion_P}
\end{equation}
where $W_{R} (k)$ is the spherical top-hat window in Fourier space and $*$ is the convolution operator.
We will consider these easily-interpretable toy expressions as conceptual references in a somewhat more realistic models of exclusion.
In these models we introduce the exclusion scale $R_{\rm{exc}}$ as a free parameter.

In reality halos are not spherical, and even for a fixed-mass sampling of the halo field the scale at which exclusion sets in (the effective “exclusion radius”) must necessarily reflect the fact that triaxiality leads to a distribution of “true” exclusion scales.
However, if one interprets halos in the context of Lagrangian density peaks, then based on the 1-D findings of
\cite{2016MNRAS.456.3985B} (where no triaxiality can be present) peak exclusion is dominated by dependence on peak height, bin width, and peak curvature.
These results appear to hold in 3-D as well \cite{2020arXiv201214404B}, so the effect of triaxiality on exclusion is likely subdominant.
Similarly, \cite{2013MNRAS.430..725V} found that triaxiality, substructure, and concentration scatter were negligible in modeling exclusion in halo two-point correlation functions.
We also find it unnecessary to model these effects for percent-level accuracy. 
The criterion used to define the halo also has an impact on exclusion, which we return to in Section \ref{section:galaxies} and Appendix \ref{appendix:halofinder}.

\subsection{Correlation function results}
\label{section:halo_cf_hm}

\begin{figure}[h!]
\center
\includegraphics[width=6.15 in, angle=0]{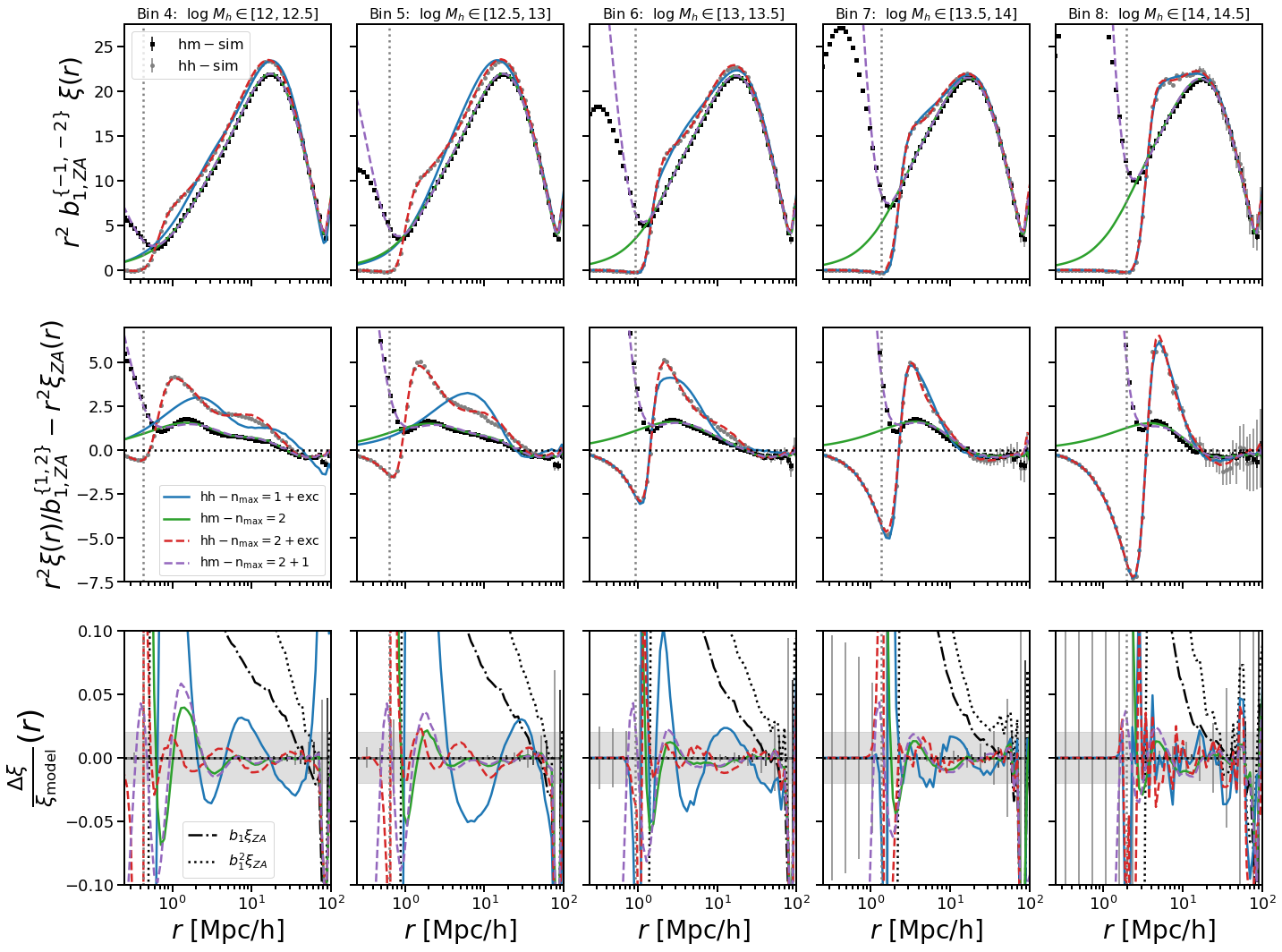}
\caption{
Fits to the halo-halo and halo-matter correlation functions in several logarithmic mass bins.
The top panels show the halo-halo and halo-matter correlation functions (multiplied by $r^{2}$ and divided by the $b_{1}^{2}$ and $b_{1}$, respectively), the center panels isolate the correction to ZA by subtracting it out, and the bottom panels show residuals with a shaded band at $2\%$.
The columns correspond to increasing halo mass from left to right.
Green (purple dashed) lines show the $\nm=2$ ($\nm=2+1$) HZPT halo-matter correlation function, while blue (red dashed) lines show the HZPT $\nm=1$ ($\nm=2$) halo-halo correlation function.
Black points show the halo-halo simulation correlation function, and grey points the halo-matter simulation correlation function.
Errors are Fourier transformed diagonal Gaussian+Poisson, which are meant as a visual guide only as errors are correlated.
The number of residual points with errorbars in the bottom panel has been reduced for visibility.
Vertical dotted lines mark the minimum scale used to fit each mass bin, which is roughly the lower limit of the transition regime for each mass bin ($\sim R_{\rm{Lag}}/2$).
\label{fig:xih_resid}}
\end{figure}

In Fig.~\ref{fig:xih_resid}, we show the correlation functions from simulations (black points for halo-matter, gray points for halo-halo) as well as various HZPT models for different mass bins (see Appendix~\ref{appendix:halos}, Table~\ref{tab:halo_table}) at $z = 0.55$.
The linearly-biased ZA (black curves) agrees well with the simulations on the largest scales considered here, but significantly underestimates the correlation function at the several percent starting at $r = 40-60~\M$.

The second row of Figure~\ref{fig:xih_resid} illustrates this deviation from linearly biased Zeldovich, which is fit by the BB terms.
There are at least three scales in the enhancement over Zeldovich in both the halo-matter and halo-halo correction functions that the BB terms must fit to account for all halo masses considered here. 
These scales are 1. a large-scale enhancement (LSE) at $\sim10~\M$ (corresponding to the $\nm=1$ parameter) 2. a small-scale enhancement (SSE) outside the halo exclusion scale (corresponding to the $\nm=2$ parameter) and 3. the halo exclusion scale.
These scales are clearly visible in the second row of panels in both halo-matter and halo-halo (though are more easily seen in halo-halo).
We will first describe how these scales vary with halo mass (as seen in row 2 of Fig.~\ref{fig:xih_resid}) and then will describe how the HZPT models explicitly account for these scales.

The enhancement in the correlation function over ZA becomes more complicated for lower mass halos.
For the largest halos ($M > 10^{13.5}~ M_{\odot}/h$, right two columns), there is only one scale or ``bump'' visible in the enhancement - the LSE and SSE coincide at several $\M$. 
Just below this unified scale is the exclusion scale, which presents itself as a vertiginous climb to profile-dominated scales in the halo-matter CF, and as a precipitous drop to zero correlation in the halo-halo CF.
For smaller halos, the single scale splits into the LSE and SSE, which are clearly visible at $\sim 5-10 ~\M$ and $\sim 1-2~\M$, respectively, for halos with $M < 10^{13}~ M_{\odot}/h$.
Physically, the SSE may be connected to the non-perturbative enhancement outside the exclusion scale observed in peak clustering observed by Ref.~\cite{2020arXiv201214404B}, while the LSE may be more related to nonlinear bias (e.g. \cite{2013JCAP...10..053V,2020arXiv201108858M}).
It would be interesting to consider an expanded hybrid modeling approach in which the LSE is modeled with a more complex nonlinear biasing model than linearly-biased ZA as the two-halo term.

We reproduce the result of H17 - halo-matter CFs are well-fit by the $\nm=2$ model above scales up to a few times the exclusion scale for all halo masses (green curve). 
This is because the $\nm=2$ model captures two scales - the LSE and SSE - quite well, and in the larger halo mass case the values of the $R_{nh}$ parameters increase and become much closer to each other, reflecting the unification of the LSE and SSE.
We attempt to slightly improve upon the $\nm=2$ model by adding a term to account for the halo profile dominance near the exclusion scale by adding a second BB term with $\nm=1$ (purple dashed curve) without compensation \footnote{We set the compensation parameter of this additional term to be very large, $10^{9}$, which is effectively the same as ignoring it.}.
This model does not show dramatic improvement over the $\nm=2$ model, but does fit the outer portion of the profile dominated region quite well.
There are small deviations in the halo-matter correlation function just outside the exclusion scale for both models - we speculate that these are the result of a too-simple treatment of halo compensation.

The halo-halo correlation function is well modeled for large masses by accounting only for the equal LSE-SSE scale and the exclusion scale through a modified $\nm=1$ model. 
To model the step in the halo (auto)correlators we model the exclusion step in a similar manner to B13 using the function $F_{\rm{exc}} (r)$ 
\begin{equation}
1
 + \xi_{hh}^{(d)} (r) =  F_{\rm{exc}} (r) \left[1+\xi_{hh}^{(c)} (r) \right],
\label{eqn:xi_d_hh}
\end{equation}
where $\xi_{hh}^{(c)}= b_{1} ~\xi_{\rm{hzpt}}$.
The function $F_{\rm{exc}} (r)$ is an approximation to a more complete physical model for exclusion \cite{2016MNRAS.456.3985B,2020arXiv201214404B} (we further discuss choices for this model in Appendix \ref{appendix:halos_exc}), and also return to it in the context of the power spectrum in Section \ref{section:trans_halo}. 
This model works quite well for the largest two halo mass bins (blue curve), but begins to fail dramatically when the LSE and SSE diverge at lower halo mass.
To address this, we take the same strategy as for the halo-matter CF and upgrade the BB term to $\nm=2$ (red dashed curve).
This results in excellent fits on scales for all halo masses in the halo-halo CF.

\subsection{Power spectrum results}
\label{section:halo_p_hm}

\begin{figure}[h!]
\center
\includegraphics[width=3 in, angle=0]{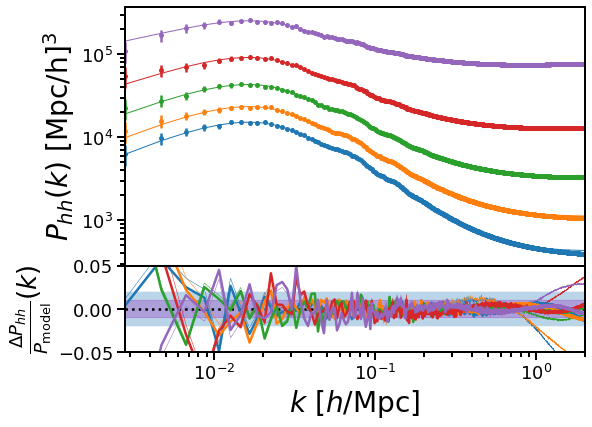}
\includegraphics[width=3 in, angle=0]{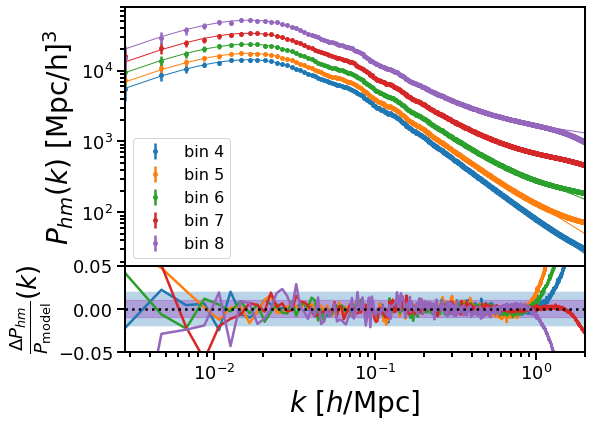}
\caption{\textit{Left:} Halo-halo power spectra residuals from and the best fit HZPT models at different halo masses.
Thin lines are residuals for the $\nm=1$ model with a free shot noise parameter, while thick lines are due to the model of exclusion with $\nm=2$ similar to the best model in Fig.~\ref{fig:xih_resid}. 
\textit{Right:} Fits for halo-matter power spectra using the $\nm=2$ model.
Colored bands are shown at 1 $\%$ and 2 $\%$.
\label{fig:Ph_resid}}
\end{figure}

In Fourier space, the $\nm=2$ model is sufficient to capture the halo-matter cross power spectrum for all halo mass bins at 2\% accuracy to $k = 1 ~\iM$, and at 1\% for almost all mass bins on the same scales (Fig.~\ref{fig:Ph_resid}).
For the halo-halo power spectrum, the $n_{\rm{max}}=1$ HZPT model in Fourier space with an added effective (not necessarily Poisson) shot noise term $\frac{1}{\bar{n}}$ appears to be accurate (thin lines in bottom panel of Fig.~\ref{fig:Ph_resid}).
In this scenario, $R_{1h}$ should be thought of as a more general $k^{2}$ term rather than as a moment of the halo profile as is the case for matter.
We discuss the small effect of removing $R$ from the auto-correlation model in Appendix \ref{appendix:halos_comp}.

The $\nm=1$ model does not explicitly account for the Fourier space effects of exclusion, but fits quite well down to $k=0.9-1 \ \iM$ at the 2\%.
For all halo masses, the scale-dependent correction due to exclusion is sub-dominant to the constant shot noise, (as seen in the left panel of Fig.~\ref{fig:Ph_resid}), and for lower halo masses the model fails to be accurate at 1-2\% at slightly lower $k$. 
This seems consistent with the results of \cite{2020MNRAS.492.5754M}, who are able to fit $P_{hh} (k)$ to lower maximum $k$ using a Lagrangian bias expansion, including a $k^{2}$ term.
However, exclusion must be properly modeled for percent-level accuracy in both configuration and Fourier space, and we return to this point in the following section, where we provide context for interpretation of the quoted accuracies in configuration and Fourier space with regard to exclusion.

As in the configuration space picture, modeling exclusion (using the Exp model presented in \ref{appendix:halos_exc}) as well as the LSE and SSE through the $\nm=2$ BB term extends the range of scales accessible to the power spectrum model. 
Proper exclusion modeling suppresses the observed deviations in the (thick) residuals to 1\% below $k=2~\iM$ and eliminates the need for a free constant shot noise parameter.
In the fits shown in the thick residuals in the bottom panel of Fig.~\ref{fig:Ph_resid}, the shot noise is fixed to Poisson, and the correction comes entirely from the exclusion model.
Using the exclusion model with an $\nm=1$ BB term suffices at the same level of accuracy for the highest two mass bins, but (as discussed in \ref{section:halo_cf_hm}) for the lower mass bins the results are worse since the SSE and LSE are distinct and must be modeled separately by something more flexible than the $\nm=1$ BB term.
While we only go to $k=2~\iM$ here, the Fourier space exclusion model is accurate to even smaller scales (see Section \ref{section:trans_halo}).

\subsection{Transforming the two-point statistics}
\label{section:trans_halo}
\begin{figure}[h!]
\center
\includegraphics[width=3 in, angle=0]{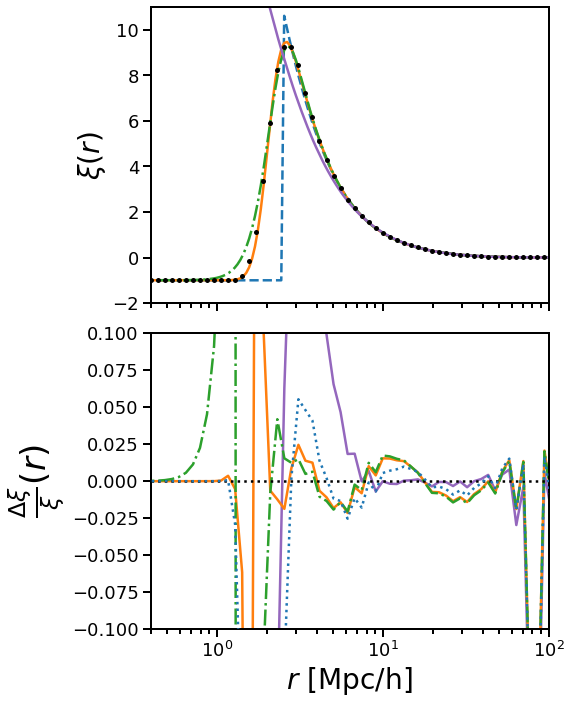}
\includegraphics[width=3 in, angle=0]{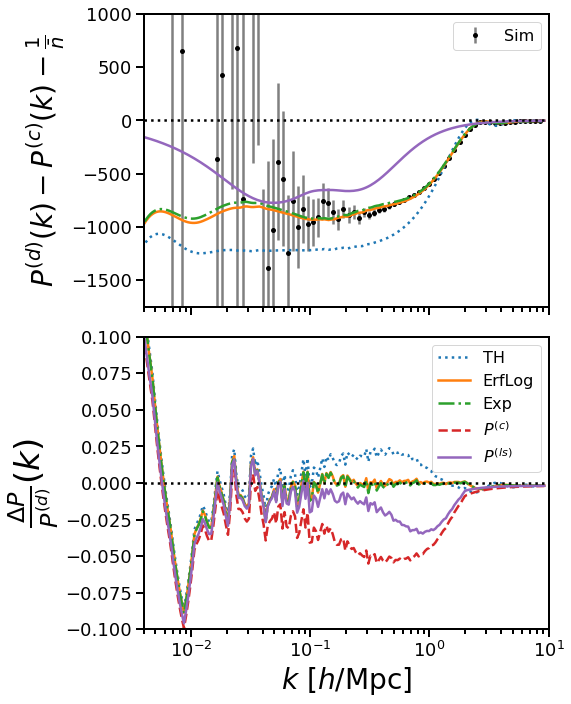}
\caption{Illustration of (HZPT) contributions to exclusion for bin 7 ($\log M_{h} \in [13.5,14]$. \textit{Left:} Correlation function with several different exclusion models. ErfLog denotes the model of B13 (with $\sigma_{\rm{exc}}$ free), while Exp denotes the model described in Appendix \ref{appendix:halos_exc} (that has correction with analytic FT).
TH denotes the simple thresholding of the continuous model, and is clearly not sufficient to capture the width of the exclusion step.
\textit{Right:} Corresponding power spectra (evaluated via FFTLog). $P^{(c)}$ denotes the (analytic) continuous model using best-fit HZPT parameters to the correlation function.
TH in this case is equivalent to the window function expression terms of B13 (but without truncating the expansion).
The solid purple curve denotes the large-scale model $P^{(ls)}$ resulting from fitting $\xi^{(ls)}$ above $10~\M$.     
\label{fig:excl_demo}}
\end{figure}

Exclusion and non-perturbative clustering present themselves differently in configuration and Fourier space.
To better understand how to interpret the accuracy of the HZPT model at the different scale cuts in $k$ and $r$, we provide a narrow comparison of models that include and exclude the non-perturbative effects of exclusion in configuration and Fourier space for a single halo mass bin (bin 7 - though the result for other bins is similar\footnote{However, for lower mass bins, the large-scale finite-size correction is super-Poisson rather than sub-Poisson, which is expected from the explanation of B13.}) in Fig.~\ref{fig:excl_demo}.

In the left panel of Fig.~\ref{fig:excl_demo}, we show two realistic models for the exclusion step in configuration space - the ErfLog model (solid orange) is the model of B13 (with free $\sigma_{\rm{exc}},R_{\rm{exc}}$, two free parameters total), and the qualitatively similar Exp model (one free parameter $R_{\rm{exc}}$, green dash-dotted) are both described in Appendix \ref{appendix:halos_exc}.
The Exp model fits the step quite well, though the shape is not quite right at the smallest scales of the step, and the ErfLog model clearly captures the exclusion step even better than the Exp model.
The ErfLog model is relatively insensitive to $\sigma_{\rm{exc}}$ - fixing $\sigma_{\rm{exc}}=0.1$ (similar to B13) only mildly degrades the accuracy of $\xi_{hh}(r)$ in the $\nm=1$ model (which is sufficient here for bin 7) with respect to the shape at the small-scale end (similar to the slight inaccuracy of the Exp model).
A benefit of the Exp form of the exclusion step is that it permits an analytic Fourier transform of $F_{\rm{exp}}(r)$ (see Appendix~\ref{appendix:halos_exc}), which explicitly displays the non-trivial $k$-dependence of the exclusion (it is not as simple as $k^{2}$) and keeps with the spirit of the HZPT model.
Clearly both of these models are capturing the correct features of non-perturbative halo clustering on the smallest scales.

For comparison, we also show the TH model (blue dashed), which is given by the simple Heaviside truncation of the $\nm=1$ HZPT model fit down to $r=2.5~\M$, as well as the same model fitted using a larger minimum scale $r=10~\M$ (solid purple), which we refer to as the "ls" (large-scale) model.
Similar to the quadratic biasing model of B13, the ls model describes the LSE but fails to capture both the full extent of the SSE outside the exclusion scale, and totally misses the exclusion step.
The un-truncated TH model ($\xi^{(c)}$) correctly describes the SSE but fails to account for the exclusion step - by adding the Heaviside truncation, the TH model provides a qualitatively correct description of both the SSE and the exclusion step.
However, quantitatively the TH model does somewhat worse than the ErfLog and Exp models by failing to account for the finite width of the step.

Fourier transforming, we look to see what happens under the various treatments of exclusion in configuration space.
The result is shown in the right panel of Fig.~\ref{fig:excl_demo}.
Here the quantity on the vertical axis (top panel) is a measure of the error induced by not treating the exclusion step and simply using the ``continuous model'' $\xi^{(c)}$ with fiducial (Poisson) shot noise $\frac{1}{\bar{n}}$.
$P^{(c)}(k)$ is the (analytic) Fourier transform of $\xi^{(c)} (r) $ that is fit up to the peak of the halo correlation function (TH in the top left panel but without the threshold).
The simulation points in the top-right panel and the red curve in the lower-right panel illustrate that modeling the SSE but totally ignoring the exclusion step induces a 5\% error on scales between $k = 0.1-1 \ \iM$, which is actually larger than if we had simply used $P^{(ls)}$, the Fourier transform of large-scale model $\xi^{(ls)}$.
This can be understood from the fact that the non-Poisson correction in the large-scale limit ($k\to0$) is given by the integral $\int d^{3}r \xi^{d}(r)$.
The large-scale model $\xi^{(ls)}$ both underestimates $\xi^{(d)}$ on scales where the SSE is relevant and overestimates it when the discrete correlation goes to zero, resulting in an accidental, but only partial, cancellation in the integral.
This results in the wrong non-Poisson correction, but one that produces a smaller residual in the bottom-right panel Fig.~\ref{fig:excl_demo} than for the continuous model $P^{(c)}$.
$P^{(c)}$ only includes positive contributions to this integral, and produces a too super-Poisson correction without the negative contributions from the exclusion step.
Adding the simple Heaviside threshold qualitatively accounts for both types of correction and tames the residuals to be about 2-3\%.
Using the more accurate exclusion step models produces negligible error at a fraction of a percent.

We note that these residuals are computed with respect to the discrete power spectrum (which includes fiducial shot noise), so all errors would be amplified were shot noise subtracted\footnote{This amplification would of course get worse at higher $k$ as the amplitude of the continuous power spectrum decreases - in practice quick calculations residuals with respect to the shot noise subtracted power spectrum for bin 6 can be above 10\% at $k=1 \ \iM$}.
Since the effect of ignoring exclusion is scale-dependent, it is not possible to cleanly relate cutoffs at a particular scale in configuration space with cutoffs at a particular wavenumber $k$ in Fourier space.
Furthermore, we see through the comparison of the residuals of $P^{(ls)}$, $P^{(c)}$, and the step models that starting on quasi-linear scales and fitting a configuration space model down to progressively smaller scales induces an exclusion correction in Fourier space that initially increases before decreasing to approach the correct non-Poisson value.
It would be interesting to further quantify how this behavior affects other biasing models that stop at a particular minimum scale $r_{\rm{min}}$ such that $R_{\rm{exc}} <r_{\rm{min}}< 10 ~\M$.

By way of this example, we see that ignoring the effects of exclusion (or stochasticity in general) as is sometimes done in small-scale biasing models of the tracer correlation functions, will necessarily result in incorrect behavior over a relatively wide range of wavenumbers in Fourier space.
Coming back to our fits to the power spectrum, this is seen in Fig.~\ref{fig:Ph_resid}, where percent-level excursions are visible in $P_{hh}$.
These excursions seem more pronounced at lower mass, reflecting the fact that the distinct LSE and SSE identified in the correlation function are not well-modeled by a single $\nm=1$ BB term.
The apparently smaller residuals for higher mass bins is, however, partially a consequence of shot noise constituting a larger fraction of the total power on small scales due to lower halo number density at higher mass.

Despite the fact that the $\nm=1$ fits in Fig.~\ref{fig:Ph_resid} don't explicitly include exclusion, they are still relatively accurate since the Fourier space BB term and free shot noise terms are implicitly modeling exclusion.
This may be explained by the fact that at leading order the $\nm=1$ BB term and the simple threshold model scale as $k^{2}$, so the BB term (or any $k^{2}$ term) might capture the leading-order behavior of exclusion.
However, a more realistic model of exclusion (e.g. the Exp or ErfLog model) is more complicated than a simple $k^{2}$ term (see Appendix~\ref{appendix:halos_exc} for the form of this expression for the Exp model).
These models introduce an additional parameter $R_{\rm{exc}}$ to the HZPT model, but knowledge of this parameter \textit{exactly gives} the value of the non-Poisson correction in the large-scale limit, eliminating the need for the free shot noise parameter in the power spectrum.
We conclude that percent-level accuracy of a model without exclusion in Fourier space is largely due to the relative importance of the free shot noise term and the leading-order $k^{2}$ behavior of exclusion, and \textit{not} due to a correct model of high-$k$ behavior, which must include the non-perturbative effects associated with halo exclusion.

\section{Galaxies and Satellites}
\label{section:galaxies}
The transition regime for galaxy-galaxy and galaxy-matter correlators is affected by both the details of satellite occupation and halo exclusion. The model of Section \ref{section:halos} for halos already includes exclusion, and we build upon that model by accounting for the presence of satellite galaxies in this section.
We consider two different galaxy samples produced according to the HOD prescriptions presented in Section \ref{section:N-body} to test the flexibility of this model.
We focus on configuration space fits and use them to estimate joint density with cosmology but also provide fits to power spectra.
We take a pragmatic approach throughout this section, using the minimal HZPT model necessary to achieve percent-level accuracy for the galaxy two-point correlators at $k\approx1~\iM$ and $r\approx1-2~\M$.

\subsection{Small-scale galaxy clustering}

The galaxy-matter correlators are analogous to the case of halo-matter correlators, but are slightly complicated by the presence of satellites.
On the largest scales considered here, galaxy-matter correlators are well-described by linear bias with ZA and compensation. 
In addition to the correlation between a particular central galaxy and the matter profile of its host halo, there is now another contribution from the correlation between matter and satellites.
The satellite fraction will impact the amplitude of the intra-halo correlations, which in turn will affect the slope and location of the transition feature described in Section \ref{section:halos}.
The smallest scale correlations are then completely governed by the halo dark matter and satellite profiles.
In our simple HOD mocks, satellites are drawn from an NFW profile.
However, since the form of the BB term is profile-agnostic due to the general form of the Pad\'e expression, there should be no great difficulty in modeling other qualitatively similar profiles (i.e. more complicated satellite profiles).

The galaxy-galaxy correlation function $\xi_{gg}$ is more complicated than the auto-correlation function for halos.
In addition to the steep drop in the correlation function near the effective exclusion scale that is expected for halos, we must consider the role of satellite galaxies.
As explored in detail by \cite{2015PhRvD..92j3516O} (see their Fig. 1) and H17, there are additional types of correlations: 1. between centrals and satellites in different halos, 2. and centrals and satellites in the same halo, as well as 3. between satellites in different halos, and 4. satellites in the same halo.
While 2. and 4. essentially serve to change the correlation function on scales relevant to the satellite profiles (i.e. roughly the combination of the profile and its self-convolution),  1. and 3. effectively introduce contributions that are versions of the central-central correlation function (with the exclusion step) that have been smoothed out over the halo scale. 
This smoothing of the exclusion scale will serve to broaden the exclusion step present for halos in the galaxy auto-correlation.

To deal with these complications, we introduce some additions to the HZPT model.
We do not model each of the terms outlined in \cite{2015PhRvD..92j3516O} separately, instead lumping some of them together into an effective HZPT model for $\xi_{gg}(r)$.
We allow for the smoothing of the exclusion step through freeing $\sigma_{\rm{exc}}$ to be larger than the value ($\approx 0.1$) that was acceptable for halos.
Additionally, we add a satellite profile term (an additional BB term with $\nm=1$) that has two free parameters (with subscript $1s$), since we fix $R_{1s}=10^{3}$.
On these scales, this choice is the same as providing no compensation for the satellite profile.
The $A_{1s}$ and $R_{1s}$ will vary depending on the details of the satellite occupation, e.g. with the amplitude scaling with the satellite fraction.
So the full equation for galaxies with exclusion (in configuration space) is:
\begin{equation}
\label{eqn:1s_xi}
\begin{aligned}
        \xi^{(\mathrm{exc})}_{gg}(r) &= \xi_{hh}(r) + b_{1} \xi^{\nm=1}_{BB,1s}(r)\\
        &= b_{1} \left[F_{\rm{exc}}\left(\xi_{ZA} + \xi^{\nm=1}_{BB} \right) + \xi^{\nm=1}_{BB,1s}(A_{1s},R_{1h,1s}) \right]
\end{aligned}
\end{equation}
where the bias $b_{1}$ is free (not fixed to the halo bias value) and we suppressed arguments except for the new parameters in the second line.
Here the 1s BB term is compatible with the usual BB interpretation and we can think of it (correctly) as a $k^{2}$ expansion in the 
satellite profile.
Adding this term does not ruin the analytic Fourier transform, which will have a form that is the product of two Lorentzians (as for matter) in Fourier space.

The effects in the transition regime for HOD mock galaxy clustering will necessarily be more complicated than that of halos (even ignoring satellites) since the HOD applies a threshold for the central occupation which spans the equivalent of several halo mass bins. 
This means that the “cross-stochasticity” (B13, \cite{2010PhRvD..82d3515H}) of exclusion in different bins will contribute more strongly to the central auto-correlations. 
However, we find that this is not something that needs to be modeled explicitly when fitting, which may have to do with the fact that the cross-stochasticity is either close to constant or of a similar scale dependence to that of the auto-stochasticity
(c.f. H17 Fig 4).
We also do not explicitly account the effects of central galaxy off-centering \cite{2012MNRAS.419.3457H,2015ApJ...806....2M}, which are relevant for an accurate treatment of the small-scale galaxy-galaxy lensing signal.
Since this effect may be accurately modeled by a modification of the profile, we anticipate that our Pad\'e term may be general enough to account for such effects.

\subsection{Two HOD mocks}

The CM and LZ samples are produced using two different underlying simulations, and complement each other in the trade-off of resolution and number of simulated cosmologies.
The CM sample is produced from a simulation with a factor of 10 larger volume than the LZ sample and allows for a cleaner test of model accuracy due to a reduction in resolution effects.\footnote{The qualitative features of the exclusion step do not appear to depend much on the use of FastPM. Using a subset of the same HOD parameters (albeit at a slightly different cosmology) we check the qualitative features (and scales) of the exclusion feature are similar using FoF catalogs produced by a TreePM code (\cite{2002ApJS..143..241W}, described in H17).}
Fits to mocks from the LZ sample have increased noise with respect to to CM mocks due to smaller volume and a resulting smaller number of HOD galaxies, but still allow us to map the HOD basis of parameters onto the HZPT basis of parameters and provide a joint distribution of the HZPT parameters and cosmological parameters.

The CM sample covers a wide range of HOD parameters that are centered on the BOSS CMASS parameter space to illustrate the flexibility of the HZPT model (parameter space described in Section \ref{section:N-body}).
The parameter space covers a wide range of satellite fractions ($f_{s} = 0.01 - 0.65$), the highest values of which are still consistent with observed galaxy samples \cite{2006MNRAS.368..715M} (though these may differ significantly from BOSS).
We emphasize that this choice of ``CMASS'' parameters is not to be taken too literally, as we do not enforce that the HOD mocks reproduce the CMASS clustering, only approximately the CMASS number density and are produced at a similar redshift.
The HOD parameter ranges are based on \cite{2014MNRAS.444..476R}, but are  taken to be more general - and use FoF rather than SO halos. 
The purpose of this sample is more illustrative and conceptual - to demonstrate that a wide range of HOD parameters can expose halo exclusion for certain mock galaxy samples, and that HZPT provides a good description of two-point correlators even in this case.
The LZ sample is more realistic in the sense that the HOD parameters are close to those favored by LOWZ clustering \cite{2020MNRAS.492.2872W,2013MNRAS.429...98P}.
These differences mean that a different minimal HZPT model is necessary to reach percent-level accuracy for each sample at the scales we address in this section.

\subsubsection{Configuration space results}

The galaxy-matter cross correlators in both the CM and LZ mock samples are well fit by the same model used for halo-matter cross correlation - the base HZPT model with $n_{max}=2$ with the linear bias $b_{1}$.
Despite the presence of satellites, the transition regime (including the outer part of the halo profile for CM) is well-modeled by the $R_{nh}$ parameters.
Fits are performed from $r=1-70 \ h/\rm{Mpc}$ in configuration space (and from $k=0.01-1 \ h/\rm{Mpc}$ in Fourier space). 
Errors for fits to the LZ power spectra include diagonal Gaussian+Poisson covariance and variance estimated from repopulating HODs at 10 different random seeds (which we take as independent of any particular HOD realization), while the CM power spectrum errors are based only on Gaussian+Poisson covariance.

\begin{figure}[h!]
\center
\includegraphics[width=3. in, angle=0]{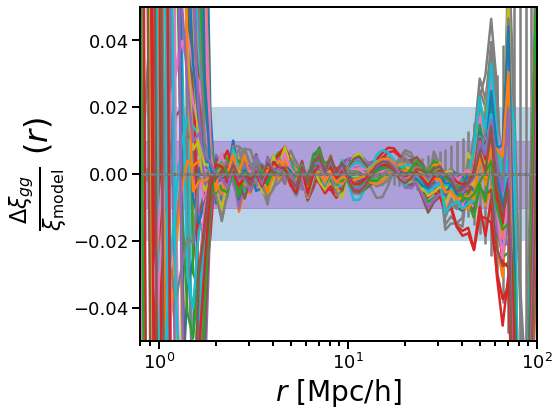}
\includegraphics[width=3. in, angle=0]{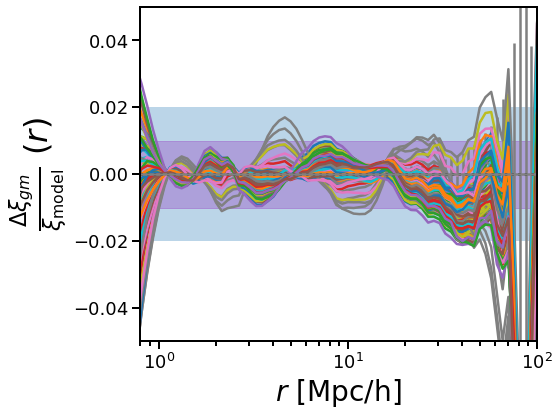}
\caption{\textit{Left}: Galaxy-galaxy correlation function residuals for the 100 CM HOD mocks.
\textit{Right}: Galaxy-matter correlation function residuals for the 100 CM HOD mocks.
Grey curves correspond to HOD mocks with $f_{s}>0.55$.
Colored regions mark 1\% (red), and 2\% (blue), errors are diagonal Gaussian+Poisson (these are meant as a visual guide only, and the errors are correlated).
The HZPT model used in these fits is the $\nm=2$ model for $\xi_{gm}$ and the satellite-enhanced $\nm=1$ model with exclusion (eqn.~\ref{eqn:1s_xi}) for $\xi_{gg}$.
\label{fig:xi_res_gal_cc}}
\end{figure}

The $\nm=2$ HZPT model is sufficient to attain several-percent accuracy in both the galaxy-matter and galaxy-galaxy correlation function (generally 1-2\%, but this is limited by uncertainty due to resolution in the case of LZ - the rms error is always less than 1\% [2\%] below $40~\M$ for $\xi_{gm}$ [$\xi_{gg}$]) down to $r = 2 \ \M $ for LZ.
For CM the $\nm=2$ model is also sufficient for 2\%-level accuracy $\xi_{gm}$ $r = 1 \ \M $, but for $\xi_{gg}$ we require the model of eqn.~\ref{eqn:1s_xi} to capture the complications due to the satellites and exclusion effects present in this sample to produce an accuracy of 2\% above $r = 2 \ \M $.
We find that for most choices of HOD parameters, fits in both gm and gg (with the satellite terms) provide fits accurate to $1\%$ down to $r\approx 0.5 \ \M$ for the CM sample.
However, to be conservative and accommodate all HOD parameters considered, here we fit $\xi_{gg}$ only down to $r=2 \ \M$.
The galaxy-matter correlation function $\xi_{gm}$ is fit to 1 \% accuracy down to $r = 1\  \rm{Mpc}/h$ for all but three of the highest satellite fraction HODs, in which case the accuracy quoted is 2\% (Fig.~\ref{fig:xi_res_gal_cc}).
All CM mock HODs with $f_{s}>0.55$ (very high for realistic LRG samples even given the general parameterization used here) have correlators plotted as gray curves in the figures.
There is a downward shift in scale in the transition regime between the two samples, which we discuss in Section \ref{section:compare-cc-ae}.

\begin{figure}[h!]
\center
\includegraphics[width=3. in, angle=0]{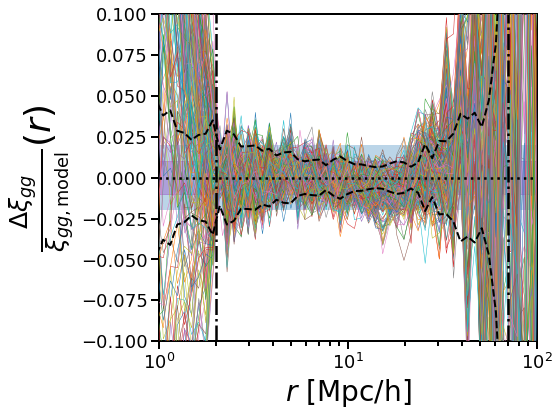}
\includegraphics[width=3. in, angle=0]{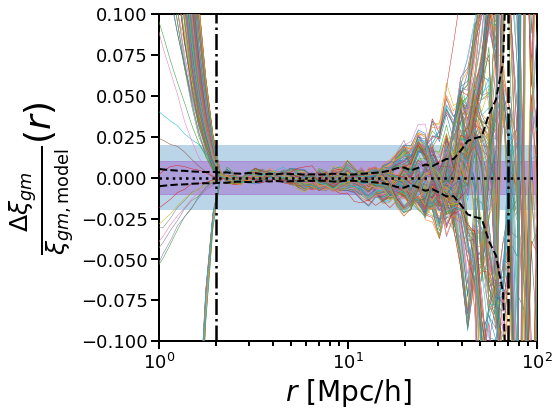}
\caption{\textit{Left:} Residuals for the $n_{\rm{max}}=2$ model for $\xi_{gg}$ for all 200 cosmology and HOD parameter combinations. (\textit{Right:}) The same for $\xi_{gm}$.
Colored bands show $1$ and $2\%$, and errors (black dashed) give $1\sigma$ Gaussian+Poisson variance (these are meant as a visual guide only, and the errors are correlated), with additional stochastic HOD error added in quadrature to the $\xi_{gg}$ error.
The HZPT model used in these fits is the $\nm=2$ model for both $\xi_{gm}$ and $\xi_{gg}$. The more involved model of eqn.~\ref{eqn:1s_xi} is not necessary for percent-level accuracy for the LZ sample.
\label{fig:xi_res_gal}}
\end{figure}

\subsubsection{Fourier space results}

\begin{figure}[h!]
\center
\includegraphics[width=3. in, angle=0]{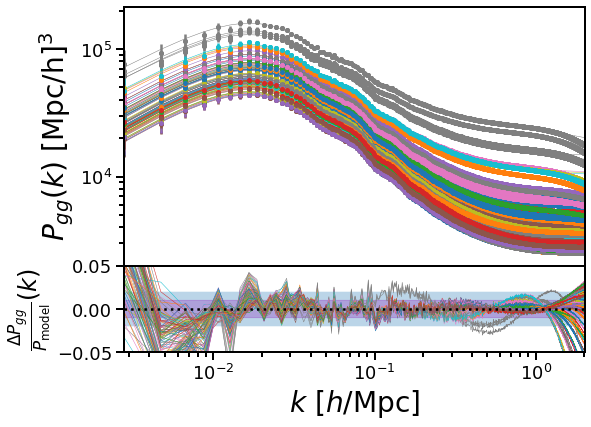}
\includegraphics[width=3. in, angle=0]{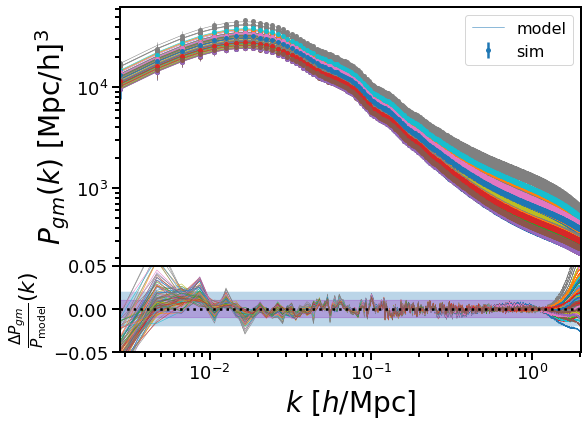}
\caption{\textit{Left}: Galaxy-galaxy power spectra and residuals for the 100 CM HOD mocks.
\textit{Right}: Galaxy-matter power spectra and residuals for the 100 CM HOD mocks.
Grey curves correspond to HOD mocks with $f_{s}>0.55$.
Colored regions mark 1\% (red), and 2\% (blue).
The HZPT model used in these fits is the $\nm=2$ model for $P_{gm}$ and the $\nm=1$ model without exclusion for $P_{gg}$.
\label{fig:P_res_gal_cc}}
\end{figure}

As discussed in Section \ref{section:halos}, exclusion in Fourier space 
is largely suppressed by scale-independent shot noise, and the $\nm=1$ model performs decently well for $P_{gg}(k)$. 
We see that for all but three of the HOD mock power spectra considered, the residuals are always less than 3\%, and are usually less than 1\%. 
Again, gray curves correspond to HOD mocks with $f_{s}>0.55$, and account for the residuals exceeding 2\%.
The three offending curves correspond to the $f_{s}>0.5$ cases mentioned for the correlation function.
From the top panel of Fig.~\ref{fig:P_res_gal_cc}, we can see that these are the highest-biased cases and the wiggle-shape of the excursions in the transition regime are what would be expected from ignoring exclusion, as we have done here. 
Contrary to the case of halos, the number density is fixed in the CM sample, so the fiducial shot noise is kept fixed as the preference for population of galaxies in halo masses changes.
For the highest mass halos, we would not expect to see the transition wiggle feature due to the high value of shot noise, but for these high satellite fraction models high-mass halos are preferred and we essentially reduce to the case of a high-mass halo bin where the exclusion feature is smoothed and shot noise is reduced (satellites act to up-weight the importance of the halo correlation with respect to the shot noise).

The fits in the power spectrum should not be taken to mean that exclusion is not important for an accurate description of two-point statistics, and the lessons of mapping between configuration space and Fourier space recounted in \ref{section:halos} still apply.

\begin{figure}[h!]
\center
\includegraphics[width=3. in, angle=0]{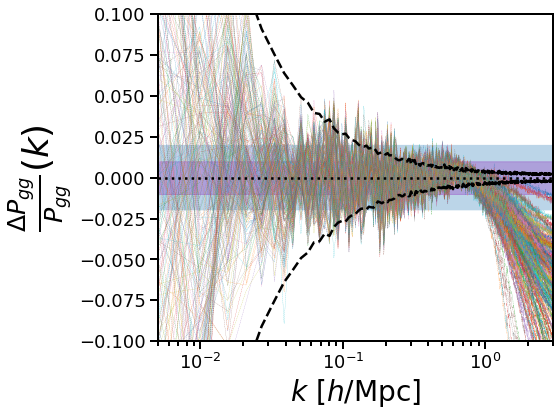}
\includegraphics[width=3. in, angle=0]{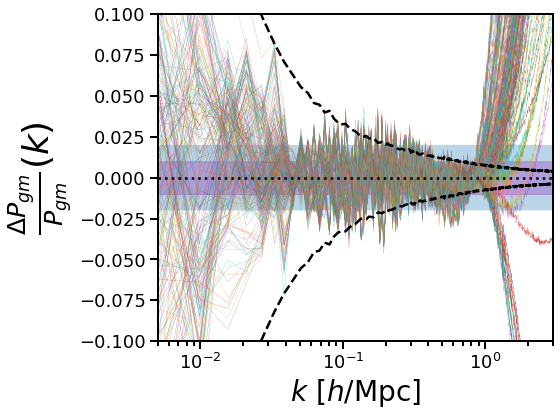}
\caption{\textit{Left:} Residuals for the $n_{\rm{max}}=1$ model for $P_{gg}$ for all 200 cosmology and HOD parameter combinations. \textit{Right:} The same for the $n_{\rm{max}}=2$ model (without exclusion) for $P_{gm}$.
Colored bands show $1$ and $2\%$, and errors (black dashed) give $1\sigma$ Gaussian+Poisson variance, with additional stochastic HOD error added in quadrature to the $P_{gg}$ error.
\label{fig:P1_res_gal}}
\end{figure}

We find that the $\nm=2$ HZPT model used for the galaxy-matter correlators is sufficient to attain several-percent accuracy (generally 2-3\%, as limited by uncertainty due to resolution - the rms error is always less than 2\% over the range of fit) down to $k = \ 1 \iM $ for the galaxy auto-correlators for the LZ sample.
For the LZ HOD mocks, the exclusion feature is essentially undetected by the fits on the scales we consider (and the feature is totally absent for ~90\% of the HOD mock correlation functions) and is not present by eye in Fig.~\ref{fig:xi_res_gal}.
Thus, the more complicated model including exclusion described above (and necessary for CM) is totally unnecessary.
This is due to the position of the exclusion feature for the LZ sample, which is at particularly small scales, and which we discuss in the next section.

We find that using an augmented two-halo term similar to the one described in Section \ref{section:ext_pmm} can result in fits that are accurate at $k\approx 10~\iM$ with a single transfer function parameter for $P_{gm}$ in some HOD realizations. 
We also find that exclusion modeling in addition to a $k$-space version of the satellite profile term similar to eqn.~\ref{eqn:1s_xi} can result in improved fits in $P_{gg}$.
An extension of the results presented in this section to higher $k$ using an augmented two-halo term and exclusion is therefore an interesting direction for future work employing HZPT modeling.

\subsubsection{Comparison of mocks}
\label{section:compare-cc-ae}

\begin{figure}[h!]
\center
\includegraphics[width=2.55 in, angle=0]{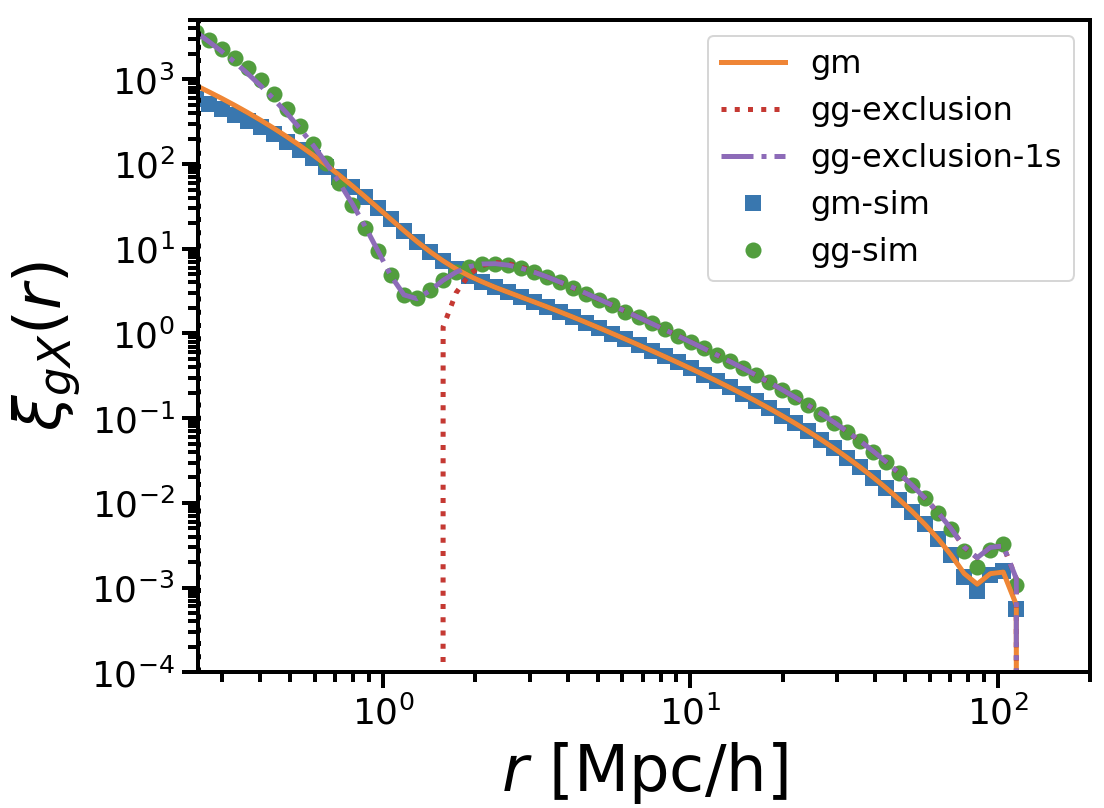}
\includegraphics[width=2.5 in, angle=0]{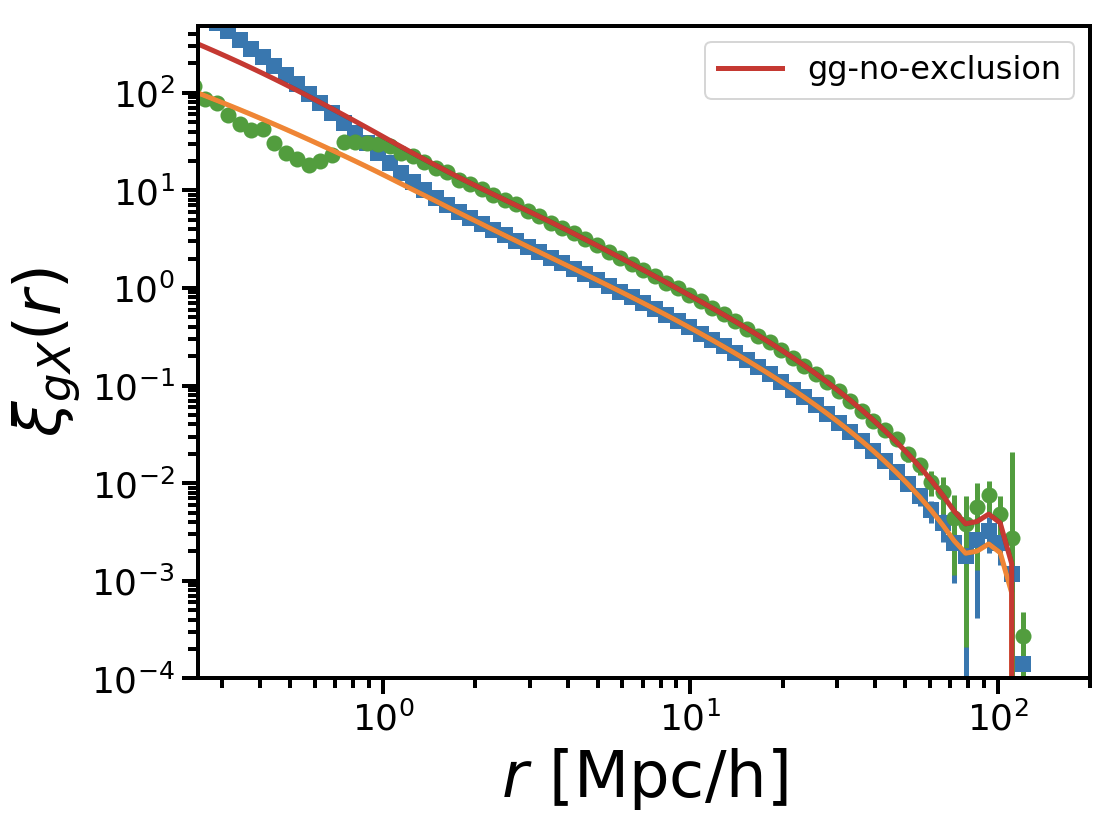}
\caption{Comparison of fits to galaxy correlators for different samples when exclusion is present. Correlation function for mock galaxies for a single HOD from the CM sample (with satellite fraction $f_{s}=0.29$).
\textit{Right:} Correlation function for mock galaxies for a single HOD from the LZ sample (with satellite fraction $f_{s}=0.01$).
\label{fig:hi_low}}
\end{figure}

Exclusion is much more prevalent in the CM sample than in the LZ sample.
When exclusion does occur, it does so on scales of $~0.6-0.7 ~\M$ in LZ\footnote{This comparison was performed using Box 20 of the Aemulus simulations, which is close to the cosmology used to generate CM, but for more disparate cosmologies the effect may be significant (already the fact that some boxes exhibit exclusion and some do not may point in this direction)} rather than around $2  ~\M$ as seems typical for CM.
For $\log M \in [13.5,14]$ (bin 7), which is completely covered by the selection of halos populated by the HOD mocks, the CM exclusion step spans $ 1.3-2.6 \ \M$, while for LZ it spans $0.7-1.0 \ \M$.
This is consistent with the scales of the dips of the exclusion features in the galaxy-galaxy correlation function for both samples (Fig.~\ref{fig:hi_low}).
For the LZ mocks, exclusion is only visible for very low satellite fractions ($f_{s}<.05$), while for the CM sample it is present the majority of the time ($\approx 80\%$).
For CM, in fact, the only time the model does not seem to show exclusion is when $\sigma_{\log M}$ is very large - suggesting that the sharp climb of the halo occupation may to some extent be driving the visibility the exclusion feature.

A reason for the difference between CM and LZ is that both the choices of halo finder and halo mass used strongly influence the exclusion feature. 
The halo catalog for LZ was produced by \texttt{ROCKSTAR} and uses ``strict SO'' 200b masses, which includes unbound particles that are not part of the \texttt{ROCKSTAR} group in a given halo.
The CM halo catalog simply uses FoF (as implemented in \texttt{nbodykit}) masses with linking length $b=0.2$.
We describe this effect in further detail in Appendix \ref{appendix:halofinder}.

Literature to date focusing on exclusion has relied on both FoF halos (e.g. B13, \cite{2020arXiv201214404B}) and SO halos (e.g. \cite{2013MNRAS.430..725V}).
If one is consistent in a choice of FoF halos or SO halos in a mock-based analysis using HOD galaxies, the treatment of exclusion will also be consistent. The LZ implementation is thus more realistic in the sense that the analysis of \cite{2014MNRAS.444..476R} used a SO halo finder to obtain the HOD constraints.
But even within the context of SO masses, 
the effect of SO vs strict SO masses can have an effect on the correlation function due to the rapid growth of the exclusion feature at small scales (Fig.~\ref{fig:halo_finder_ab}).
In HOD-based modeling it is then desirable to in some sense marginalize over halo definition.
Ref. \cite{2020MNRAS.492.2872W} used a free parameter ($R_{\rm{rescale}}$) to do this, using the justification that the main effect of varying halo definition is the effect on halo radii as related to matter and satellite profiles.
However, as shown here, the impact of halo definition on tracer correlators is not just through the one-halo term, but also through the two-halo term via exclusion.

While in Section~\ref{section:halos}, finding the minimum effective HZPT model that is accurate at the percent level required careful modeling of non-perturbative halo clustering, for mock HOD galaxies we find the reality to be more complicated.
Depending on the galaxy sample and host halo definition, exclusion may or may not be an effect that is necessary to model at this level of accuracy. 
The CM sample contains HOD populations that often require a more involved model of halo exclusion and satellite contributions, while the LZ sample only shows exclusion on scales smaller than $1~\M$, so we can safely ignore them when aiming for percent-level accuracy above these scales.

\subsection{Correlations with cosmological parameters}
\label{section:ae_cosmo}

\begin{figure}[h!]
\center
\includegraphics[width=5.5 in, angle=0]{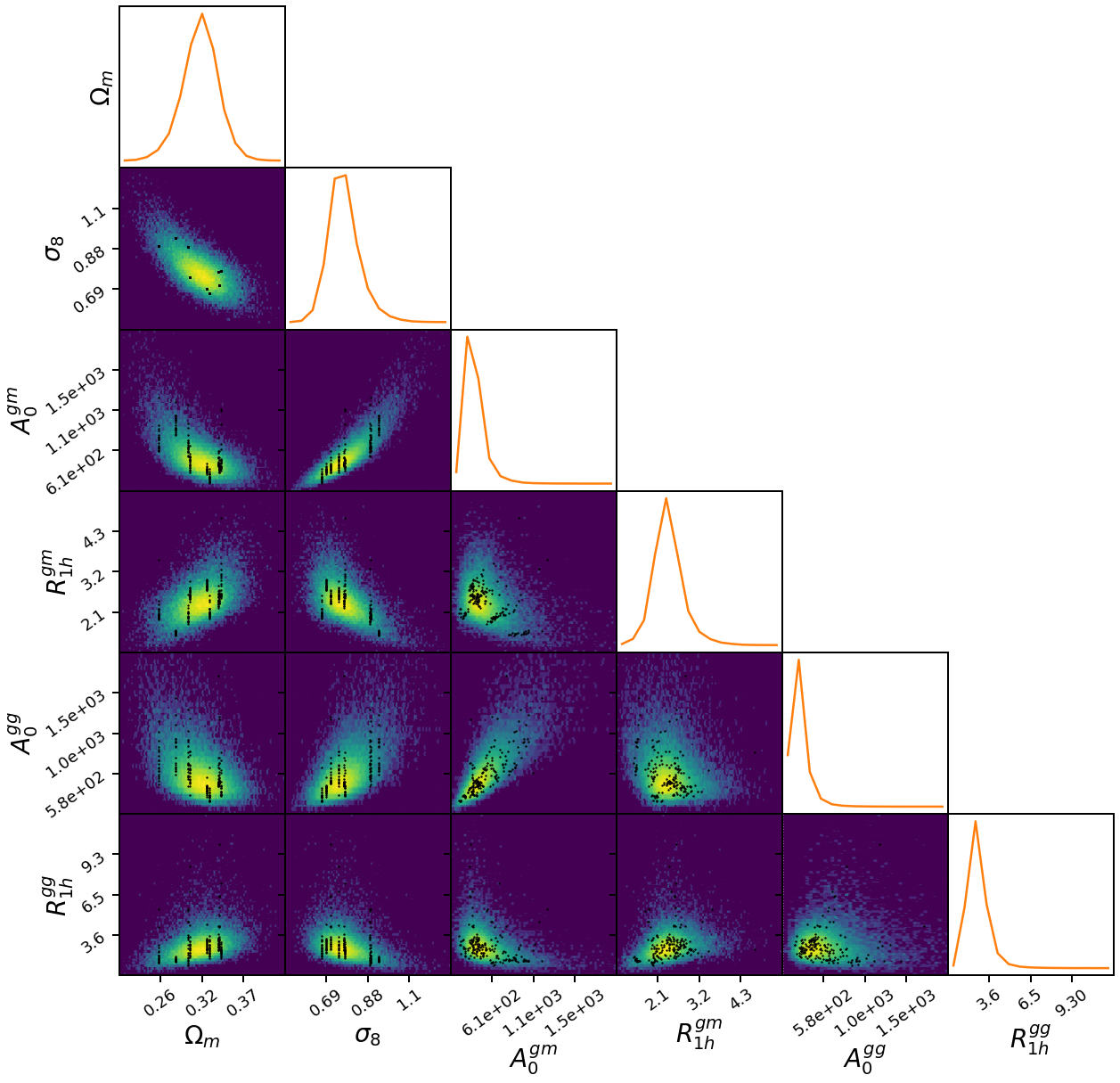}
\caption{Projection of density estimated from the best-fit HZPT parameters for the LZ sample using 10 of the Aemulus simulations ($\Omega_{m}, \sigma_{8}$).
Best-fit parameters used for training are overplotted as black points.
The $R$, $R_{1}^{2}$ and $R_{2h}$ parameters also fitted are not shown for visibility, and are often strongly degenerate with the displayed parameters.
\label{fig:density}}
\end{figure}

We use the LZ sample and the best-fit HZPT parameters for galaxy-matter and galaxy-galaxy two-point correlators to outline an approach to estimating a joint prior of HZPT model parameters and cosmological parameters.
In particular, we use Sliced Iterative Generator (SIG) \cite{2020arXiv200700674D} to perform density estimation using the best-fit HZPT parameters obtained from the set of 200 LZ correlation functions as training points. 
Potential analyses employing HZPT as the model for two-point correlators could then use such a density as a prior for analysis.
Fig.~\ref{fig:density} shows projections of a reduced density (considering only $\Omega_{m}$ and $\sigma_{8}$) for HZPT parameters that vary significantly with $\Omega_{m}$ and $\sigma_{8}$.
The fitted $R, R_{1}^{2},R_{2h}$ parameters not shown do vary significantly with $\Omega_{m}$ and $\sigma_{8}$, but have strong degeneracies with $R_{1h}$ so have been removed for visibility.
We also note that, as mentioned in the previous section, we are usually not fitting non-perturbative effects near halo scales in the LZ sample. 

This approach trades interpolation error (as in a standard emulator) for approximation error (through density estimation) and is more flexible than a typical emulator, as one can tweak the priors manually without running a new set of expensive simulations.
Of course, there should be a physical reason for shifting or narrowing the priors, but simply broadening the priors may account for increased uncertainty about a particular galaxy or tracer sample.
We also anticipate that constructing such an emulator-style tool from HZPT would require fewer training samples evaluated at distinct values of cosmological parameters.
This is because the forms of the HZPT correlators are more restrictive than the form of a typical emulator. 

The galaxy-galaxy and galaxy-matter $A_{0}$ parameters show a negative correlation with $\Omega_{m}$ and a positive correlation with $\sigma_{8}$, both of which are consistent with the sign of the power laws fitted for the matter power spectrum in Section~\ref{section:matter_baryon} and in SV15 (for $\sigma_{8}$).
The parameter $R_{1h}$ seems to have the opposite relationship, which might at first appear surprising. 
However, note the relationship between cosmological parameters shown in the top left panel.
This is a reflection of the design strategy of the Aemulus simulations, which are based on Planck constraints \cite{2019ApJ...875...69D}.
Since $\Omega_{m}$ increases as $\sigma_{8}$ decreases, it is not straightforward to meaningfully disentangle dependence on $\Omega_{m}$ and $\sigma_{8}$ individually. 
In any event, it is clear that the HZPT parameters shown are not independent of cosmology, and depend strongly on at least some combination of $\Omega_{m}$ and $\sigma_{8}$ with scatter that is captured in the density width.
We aim to provide a more quantitative comparison of the effect of different HZPT priors on cosmological constraining power in future work.

\section{Conclusions}
\label{section:conclusions}
In this paper we expand the Halo-Zeldovich Perturbation Theory approach to accurately model two-point correlators of matter and tracers well into the nonlinear regime.
For all correlators, the models are generally accurate at the percent level down to $k<1~\iM$ ($r>2~\M)$ in Fourier (configuration) space.
A summary of HZPT models used in this paper is provided in Table~\ref{tab:model_table}.
An additional benefit of two-point correlators in HZPT is that the corrections to the (linearly biased) ZA contributions have analytic Fourier transforms, and we provide expressions and fits for both forms of the correlators.
Being analytic, this model is well suited to fast inference and gradient-based sampling, and we make the model and gradients available through a lightweight \texttt{python} package\footnote{\url{https://pypi.org/project/gzpt/} \href{https://github.com/jmsull/gzpt}{\textsc{\faGithub}}}.

\begin{table}[h!]
    \centering
    \begin{tabular}{|c|c|c|}
    \hline  
        Model & Parameters & $(k_{\rm{max}},~r_{\rm{min}})$ \\
        \hline 
        $mm : \nm=2$ & $\{A_{0}, R, R_{1h},R_{1}, R_{2h}\}$ &  $(1,1)$   \\
        $mm : \nm=3$ & $\{A_{0}, R, R_{1h},R_{1}, R_{2h},R_{2}, R_{3h}\}$ &  $(8,\cdot)$  \\
        $mm : \mathrm{alt} + \nm=2$ & $\{A_{0}, R, R_{1h},R_{1}, R_{2h},\alpha, \beta \}$ &  $(10,\cdot)$  \\
        \hline
        $hm : \nm=2$ & $\{b_{1}, A_{0}, R, R_{1h},R_{1}, R_{2h}\}$ &   $(1,2)$    \\
        $hh : \nm=1  \  $ & $\{\frac{1}{\bar{n}_{\rm{eff}}},  b_{1}, A_{0}, R, R_{1h}, \}$ &  $(1,\cdot)$ \\
        $hh : \nm=2+\rm{exc}  \  $ & $\{  b_{1}, A_{0}, R, R_{1h}, R_{1}, R_{2h}, R_{\rm{exc}}, (\sigma_{\rm{exc}})\}$ &  \ $(2,2)$  \\
        \hline
        $gm : \nm=2$ & $\{b_{1}, A_{0}, R, R_{1h},R_{1}, R_{2h}\}$ &  $(1,2)$ \\
        $gg : \nm=1  \  $ & $\{\frac{1}{\bar{n}_{\rm{eff}}},  b_{1}, A_{0}, R, R_{1h}, \}$ &  $(1,\cdot)$ \\
        $gg:  \nm = 2  \  $ & $\{ b_{1}, A_{0}, R, R_{1h},R_{1}, R_{2h}\}$ &  $(\cdot,2)$  \\
        $gg:  \nm=1+\rm{exc}+1s  \  $ & $\{ b_{1}, A_{0}, R, R_{1h}, R_{\rm{exc}}, \sigma_{\rm{exc}}, A_{1s}, R_{1s,1h}\}$ & $(\cdot,2)$ \\
        \hline  
    \end{tabular}
    \caption{A subset of HZPT models used in this paper and their ranges of validity for different correlators. 
    Free shot noise constants $\frac{1}{\bar{n}_{\rm{eff}}}$ are only applicable for power spectra.
    Scales are quoted in units of $(~[\iM], ~[\M])$ in the last column when available.
    The accuracy of all models in this table is \textit{at least} 2\%, but see individual sections for details. }
    \label{tab:model_table}
\end{table}

We demonstrate that the effect of a wide range of baryonic feedback models on the matter power spectrum - as implemented by hydrodynamical simulations - can be accounted for within the HZPT framework.
These changes can be understood in terms of the halo model (see Appendix~\ref{appendix:baryonification}).
We also provide two extended models extending the one and two-halo terms.
The extended two-halo term improves upon the ZA and can reach 1\% accuracy out to $k=10~\iM$ when paired with the $\nm=2$ BB term for dark matter.
The extended one-halo term model is of comparable accuracy to contemporary nonlinear models (Appendix~\ref{appendix:hmcode}) out to $k \approx 8 ~\iM$ for matter (including a high-feedback model of baryonic effects) and we provide power law scalings to account for variation in dark matter correlators with respect to the cosmological parameters $\Omega_{m}$ and $\sigma_{8}$.

Halo-matter and halo-halo correlators are well-described by the HZPT model when the non-perturbative phenomenon of halo exclusion is accounted for.
Halo clustering is characterized in configuration space by a large-scale enhancement above $\sim10~\iM$, a non-perturbative small-scale enhancement above the exclusion scale, and a sharp step at the exclusion scale.
We provide a one-parameter analytic model for the exclusion step that has analytic Fourier transform, and the small-scale enhancement is well-modeled by the BB correction term.
Properly modeling halo exclusion in configuration space guarantees the correct behavior in Fourier space, including exactly accounting for sub/super-Poisson shot noise in the large-scale limit.
Including shot noise in the residuals, the amplitude of the exclusion contribution with respect to the total halo-halo power spectrum is relevant at the several percent level for $k>0.1~\iM$ and the contribution is scale-dependent.
We emphasize that perturbative models of halo bias that attempt to describe small scales are necessarily incomplete without a (non-perturbative) model of halo exclusion, and will fail dramatically in configuration space near the exclusion scale.
We find that without explicitly modeling exclusion in the power spectrum, we obtain residuals less than 2\% below $k\approx 0.7~\iM$, as the $\nm=1$ BB term along with a free constant shot noise appears to account for the leading-order effect of exclusion.
However, we warn that perturbative models of halo clustering that appear accurate in Fourier space at these higher $k$ are in part modeling the scale-dependent effects of non-perturbative exclusion.

Galaxy-galaxy and galaxy-matter correlators are accurately captured by HZPT in the context of LRG-like HOD mock galaxies.
Exclusion can be relevant in the simulated galaxy-galaxy correlation function for certain choices of HOD parameters, halo mass, and halo finder.
We provide an estimated density that captures the relation of HZPT parameters with $\Omega_{m}$ and $\sigma_{8}$ for HOD mocks that closely resemble the BOSS LOWZ sample.
Recently, emulators for tracer two-point statistics have become extremely popular \cite{2019ApJ...874...95Z,2015ApJ...810...35K,2019ApJ...884...29N,2021arXiv210111014K,2019MNRAS.484..989W,2021arXiv210100113M,2019JCAP...02..031R, 2020MNRAS.492.2872W, 2020PhRvD.102f3504K, 2019JCAP...12..057V}, and have provided useful and effective interpolations of simulation statistics.
However, these surrogate models are usually complicated to construct and are often dependent on a number of hyperparameters, making them opaque to interpretation even beyond the inability to write down a simulation as a closed-form model.
One can easily use HZPT to build a more interpretable tool similar to an emulator through the simple approach outlined in Section \ref{section:ae_cosmo} using estimated priors to quantify uncertainty.

For certain galaxy samples, it is possible that exclusion may stand out in the observed projected correlation function (depending on satellite fraction, host halo mass, or selection effects - there is perhaps a hint of this in right panel of Fig. 3 of \cite{2020arXiv201001143Z}).
Whether or not halo exclusion is important to include in an effective model for use in analyses is at the very least an assumption that should be checked, especially as more diverse tracers become widely used in future surveys.

One aspect of tracer auto-correlations we did not treat in this paper is the effect of cross-stochasticity, which might be especially relevant for combining populations of tracers occupying significantly different mass halos.
We did not explore the redshift dependence of the model parameters for tracers, but anticipate it may be fit relatively simply as in H17.
We also have not treated redshift-space distortions or other observational systematics, which of course are essential for connecting to observed two-point statistics.
Finally, assuming Lagrangian density peaks are the sites of halo formation, halo exclusion will also depend on cosmology at some level \cite{2016MNRAS.456.3985B}, and so is of particular interest as measurements of cosmological information through large-scale two-point statistics of LSS are saturated and small scales remain potentially under-extracted.

HZPT serves to bring analytic descriptions of two-point correlators further into the transition and nonlinear regimes, and is a fast and interpretable complement to simulation-based models.
The success of HZPT on small scales illustrates the flexibility of Pad\'e-type expressions for modeling two-point correlators.
The form of these expressions is quite simple compared to multi-loop PT, modified halo and HOD models, and most emulators.
We expect that the small-scale treatment of so-called ``3x2pt'' analyses can be significantly improved by leveraging HZPT.

\acknowledgments
We use \texttt{numpy} \cite{2011CSE....13b..22V}, \texttt{scipy} \cite{2020NatMe..17..261V}, \texttt{astropy} \cite{2018AJ....156..123A}, \texttt{nbodykit} \cite{2018AJ....156..160H}, \texttt{corrfunc} \cite{2020MNRAS.491.3022S}, \texttt{halotools} \cite{2017AJ....154..190H} and \texttt{FFTW} \cite{2012ascl.soft01015F}, \texttt{CLASS} \cite{2011JCAP...07..034B}, \texttt{CAMB} 
\cite{2000ApJ...538..473L}, and \texttt{mcfit} \cite{2019ascl.soft06017L}.
We thank Martin White, Shi-Fan Chen, Zvonimir Vlah, and Joseph DeRose for helpful comments on a draft version of this paper.
We also thank Shi-Fan Chen for sharing an early version of fast Zeldovich code from \href{https://github.com/sfschen/velocileptors}{\texttt{velocileptors}}, Zvonimir Vlah for providing a notebook used to compute integrals for \cite{2017JCAP...10..009H} as well as suggesting Section \ref{sec:alt_two_halo}, and Joseph DeRose for assistance with the Aemulus simulations.

JMS acknowledges support from the U.S. Department of Energy Computational Science Graduate Fellowship. 
SS acknowledges support from the McWilliams postdoctoral fellowship at Carnegie Mellon University.
This material is based upon work supported by the National Science Foundation under Grant Numbers 1814370 and NSF 1839217, and by NASA under Grant Number 80NSSC18K1274.

This research used resources of the National Energy Research Scientific Computing Center (NERSC), a U.S. Department of Energy Office of Science User Facility located at Lawrence Berkeley National Laboratory, operated under Contract No. DE-AC02-05CH11231.
This research has made use of NASA's Astrophysics Data System.

This material is based upon work supported by the U.S. Department of Energy, Office of Science, Office of Advanced Scientific Computing Research, Department of Energy Computational Science Graduate Fellowship under Award Number DE-SC0019323.
This report was prepared as an account of work sponsored by an agency of the United
States Government. Neither the United States Government nor any agency thereof, nor any of their employees, makes any warranty, express or implied, or assumes any legal liability or responsibility for the accuracy, completeness, or usefulness of any information, apparatus, product, or process disclosed, or represents that its use would not infringe privately owned rights. Reference herein to any specific commercial product, process, or service by trade name, trademark, manufacturer, or otherwise does not necessarily constitute or imply its endorsement, recommendation, or favoring by the United States Government or any agency thereof. The views and opinions of authors expressed herein do not necessarily state or reflect those of the United States Government or any agency thereof.

\appendix
\section{Details of Halo-Zeldovich Perturbation Theory}
\label{appendix:theory}

\subsection{Full Model ($n_{\rm{max}} \leq 2$)}
\label{appendix:theory_model}
Here we provide the full HZPT model in its basic form, with explicit expressions for $\nm = 0,1,2$.
\begin{center}
\begin{align}
    P^{mm} = P_{\rm{Zel}} + P_{BB}^{mm} \\
    P^{tm} = b_{tm} (P_{\rm{Zel}} + P_{BB}^{tm}) \\
    P^{tt} = \frac{1}{\bar{n}_{t}} + b_{tt}^{2} (P_{\rm{Zel}} + P_{BB}^{tt}) \\
    P_{BB} = A_ {0} \left(1- \frac{1}{1+k^{2}R^{2}} \right) \frac{
    \sum\limits_{n=0}^{n_{\rm{max}} -1}
    k^{2n}R_{n}^{2n}}{
    \sum\limits_{n=0}^{n_{\rm{max}}}
    k^{2n}R_{nh}^{2n}}
\end{align}
\end{center}

\noindent
$n_{\rm{max}}=0$
\begin{centering}
\begin{align}
    P_{BB} (k) =   F_{\rm{comp}}(k) = A_{0} \left(1- \frac{1}{1+k^{2}R^{2}} \right)\\
    \xi_{BB}(r) = F_{\rm{comp}} (r) = - A_{0} \frac{e^{-\frac{r}{R}}}{4 \pi R^{2}}
\end{align}
\noindent
$n_{\rm{max}}=1$
\begin{align}
    P_{BB}(k) =  P_{BB} = F_{\rm{comp}}(k) \frac{1}{1 + k^{2} R_{1h}^{2}} \\
    \xi_{BB}(r) = F_{\rm{comp}}(r) \frac{\left(
    1-\left( \frac{R}{R_{1h}} \right)^{2} e^{ - \frac{R - R_{1h}}{R R_{1h}} r } \right)}{\left(1-\left(\frac{R_{1h}}{R}\right)^{2}\right)}
\end{align}
\end{centering}

$n_{\rm{max}}=2$
\begin{centering}
\begin{align}
    P_{BB}(k)  = F_{\rm{comp}}(k) \frac{1 + k^{2}R_{1}^{2}}{1 + k^{2} R_{1h}^{2} + k^{4}R_{2h}^{4}} \\
    \xi_{BB}(r) = F_{\rm{comp}}(r) \left(\frac{\left( 1-(\frac{R_{1}}{R})^{2}\right) + A e^{r\left(\frac{1}{R} -\frac{\sqrt{R_{1h}^{2} -S}}{\sqrt{2} R_{2h}^{2}} \right)} + B e^{r\left(\frac{1}{R} -\frac{\sqrt{R_{1h}^{2} + S}}{\sqrt{2} R_{2h}^{2}} \right)}  }{\left(1 - (\frac{R_{1h}}{R})^{2} -(\frac{R_{2h}}{R})^{4} \right)}  \right),\\
    S = \sqrt{R_{1h}^{4} -4 R_{2h}^{4}}, \\
    A = \frac{R^{2} (-2 R_{2h}^{4} + R1^{2} (R_{1h}^{2}-S) ) + R_{2h}^{4} (R_{1h}^{2}-S) + R_{1}^{2} (-R_{1h}^{4} + 2 R_{2h}^{4} + R_{1h}^{2} S)}{2 S R_{2h}^{4}}, \\
    B = - \frac{R^{2} (-2 R_{2h}^{4} + R1^{2} (R_{1h}^{2}+S) ) + R_{2h}^{4} (R_{1h}^{2}+S) + R_{1}^{2} (-R_{1h}^{4} + 2 R_{2h}^{4} - R_{1h}^{2} S)}{2 S R_{2h}^{4}}.
\end{align}
\end{centering}

In this work, we reparameterize $R_{2h}$ as $R_{2h} \equiv \frac{ R_{1h}}{\sqrt{2}R_{12}}$ and vary the parameter $R_{12}$ in our fits.
We enforce $R_{12}\geq1$, since otherwise the $\nm=2$ correlation function BB term takes on imaginary values.
Analytic gradients in the python package are adjusted accordingly to be gradients of $R_{12}$.

\begin{figure}[h!]
\center
\includegraphics[width=3. in, angle=0]{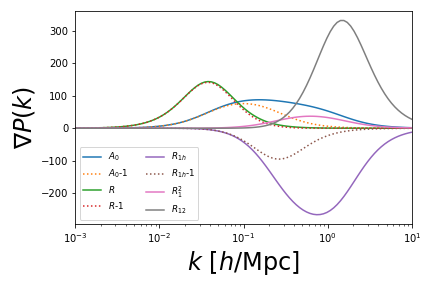}
\includegraphics[width=3. in, angle=0]{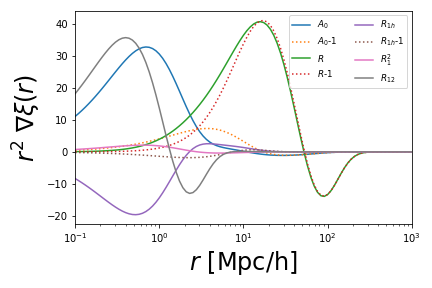}
\caption{Gradients of the $P_{BB}(k)$ (\textit{left}) and $r^{2}~\xi_{BB}(r)$ (\textit{right}).
Solid lines are gradients for the $\nm=2$ model, while dashed lines are for the $\nm=1$ model.
For visibility for $P_{BB}$ the $A_{0}$ and $R$ gradients have been multiplied by $100$ and $10$, respectively.
For visibility for $\xi_{BB}$ the $A_{0}$ and $R$ gradients have both been multiplied by $1000$.
\label{fig:gradients}}
\end{figure}

We show the BB term gradients for $\nm=1,2$ in Fig.~\ref{fig:gradients}.

\subsection{Profile expansion}
\label{appendix:theory_profile}
This closely follows Section 2 of MS14.
\noindent
Starting from \ref{eqn:kprofile}, we expand the $j_{0}(kr)$ integral in its argument:
\begin{equation}
\label{eqn:appA2_uM}
    u_{M}(k) = \frac{4 \pi}{M} \int_{0}^{R_{\rm{halo}}} dr~r^{2}~\rho_{M} (r)
    \left( 1 - \frac{k^{2} r^{2}}{3!} + \frac{k^{4} r^{4}}{5!} - ...\right)
\end{equation}
\noindent
and the modulus squared of the profile is
\begin{equation}
\label{eqn:appA2_uMexp}
    |u_{M}(k)|^{2} = |\mathcal{F}_{0}(M)k^{0} - \mathcal{F}_{1}(M)k^{2} + \mathcal{F}_{2}(M)k^{4} - ...|^{2} 
\end{equation}
\noindent
where
\begin{equation}
\label{eqn:appA2_Fn}
    \mathcal{F}_{n}(M) \equiv \frac{4 \pi}{(2n+1)! M} \int_{0}^{R_{\rm{halo}}} dr~r^{2(1+n)}~\rho_{M} (r).
\end{equation}

\noindent Then the one-halo term becomes 
\begin{equation}
\label{eqn:appA2_full_exp}
    P_{1h}(k) = \int dn(M) \frac{M}{\bar{\rho}} \left(\mathcal{F}_{0}^{2} k^{0} -2\mathcal{F}_{0}\mathcal{F}_{1} k^{2} + (\mathcal{F}_{1}^{2} + 2 \mathcal{F}_{0}\mathcal{F}_{2}) k^{4} - ...\right), 
\end{equation}
\noindent
which, for order $n_{\rm{max}}$, has Pad\'e approximant with the same roots as given by the BB term in the previous section.

\noindent
Introducing clarifying notation for the first few terms
\begin{equation}
\label{eqn:appA2_final_exp}
    P_{1h}(k) = A_{0}(1 - \widetilde{R}_{1}^{2} k^{2} + \widetilde{R}_{2}^{4} k^{4} + ...),
\end{equation}
where $A_{0}$ is the same as in the BB term.
Explicitly, for $n_{\rm{max}}=1$, the relation between the BB parameters and the profile expansion parameters is simply $R_{1h} = \widetilde{R}_{1}$, while for $n_{\rm{max}}=2$ the relations are the following:
\begin{equation}
\label{eqn:appA2_R1h}
    R_{1h}^{2} = \left( \frac{\widetilde{R}_{1}^{2} \widetilde{R}_{2}^{4}}{\widetilde{R}_{1}^{4} -\widetilde{R}_{2}^{4}} \right),
\end{equation}

\begin{equation}
\label{eqn:appA2_R2h}
    R_{2h}^{4} = \left( \frac{ \widetilde{R}_{2}^{8}}{\widetilde{R}_{1}^{4} -\widetilde{R}_{2}^{4}} \right)
\end{equation},

\begin{equation}
\label{eqn:appA2_R1sq}
    R_{1}^{2} = \left( \frac{2 \widetilde{R}_{1}^{2} \widetilde{R}_{2}^{4} - \widetilde{R}_{1}^{6}}{\widetilde{R}_{1}^{4} -\widetilde{R}_{2}^{4}} \right).
\end{equation}

\noindent
For $R_{1}^{2}\geq0$, we have $\widetilde{R}_{1} \geq \widetilde{R}_{2} \geq  \frac{\widetilde{R}_{1}}{\sqrt{2}}$

\subsection{Deriving $R_{n,nh}$ from the halo model with baryonic effects}
\label{appendix:baryonification}

To illustrate that it is possible to derive the HZPT $R_{n,nh}$ profile parameters in the context of baryonic effects from the halo model, we calculate them for the simple case of the mass function of \cite{2008ApJ...688..709T} (M200c) and NFW profile \cite{1996ApJ...462..563N} (with concentration $c=5$ for simplicity).
We approximate the impact of baryons on the HZPT parameters via  ``baryonified'' profiles as modeled in \cite{2020JCAP...04..019S} (and constrained by X-ray data).
These profiles incorporate the presence of stellar mass and satellite galaxies, gas that has been pushed by feedback toward the edge of the halo, and the resulting response of the dark-matter profile to these changes (see Section 2 of \cite{2020JCAP...04..019S}).
We fix the free parameters of the baryon-matter profile to the best-fit values of \cite{2020JCAP...04..019S} corresponding to the ``Model B-avrg'' scenario: $\eta_{\rm{cga}}=0.6$, $\eta_{\rm{star}}=0.32$ for the stellar profile, and $M_{c}=6.6 \times 10^{13} M_{\odot}/h$, $\mu=0.21$, $\theta_{\rm{ej}}=4$ in the determination of the gas profile.

Performing the integrals of eqn. \ref{eqn:appA2_Fn}, we find, for the dark matter only (dmo) profile, that  $(A_{0}$,$R_{1h}$,$R_{1}^{2}$,$R_{2h})_{\rm{dmo}} =$ $ (1114~[\iM]^{3}$, $6.1~\M$, $35 ~[\M]^{2}$, $2.9~\M)$, and, for the baryon and dark matter profile (dmb), that  $(A_{0}$,$R_{1h}$,$R_{1}^{2}$,$R_{2h})_{\rm{bdm}} =$  ($1112~[\iM]^{3}$, $5.1~\M$, $24~[\M]^{2}$, $2.6~\M$).

The difference in $A_{0}$ is negligible, as we would expect since $A_{0}$ is essentially the one-halo amplitude (which was fixed in Section \ref{section:baryons}  along with $R$).
However, the relative changes in the $R_{n,nh}$ parameters are significant, and are similar to those discussed in Section \ref{section:baryons} (though the correspondence is not exact given the idealized setting).
Here we may clearly identify the source of the change in parameters (after translating mass-integrated profile moments $\tilde{R}_{n}$ to Pad\'e parameters $R_{n,nh}$). 
The $\mathcal{F}_{1}(M)$ and $\mathcal{F}_{2}(M)$ values both grow uniformly in mass with respect to the dmo case, since in the dmb case the gas is pushed out toward the outskirts of the halo, where matter contributes more to the $r^{4},~r^{6}$ moments.

The radius at which to truncate the $\mathcal{F}_{n}$ integrals is somewhat unclear, but here we use a scale close to the truncation radius of the NFW profile for definiteness (with $R_{\rm{halo}}=\frac{9}{8}\epsilon r_{200c}$ with $\epsilon=4$).
This choice of truncation radius results in good agreement in the enclosed masses of the dmo and dmb profiles, and is small enough to prevent the integrals from being sensitive to two-halo contributions (as modeled in \cite{2020JCAP...04..019S}).
There is a very small difference ($\sim0.1$\%) in enclosed masses that is the source of the negligible difference in $A_{0}$ for the two profiles. 
Given the uncertain nature of realistic matter profiles (e.g. scatter in the concentration-mass relation), it remains advantageous to take an agnostic attitude toward halo profile details and avoid integrating directly as we do in the main text.

\subsection{Comparison to \texttt{HMCode2020}}
\label{appendix:hmcode}
\begin{figure}[h!]
\center
\includegraphics[width=4 in, angle=0]{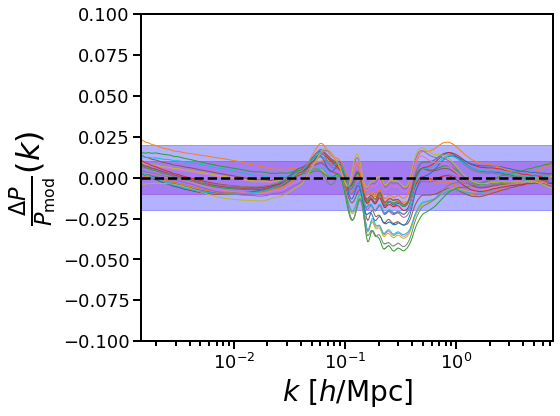}
\caption{Residuals for \texttt{HMCode2020} with respect to the same CosmicEmu test set shown in the right panel of Figure~\ref{fig:matter_nmax3_cosmo_test_train}.
\label{fig:hmcode}}
\end{figure}

Figure~\ref{fig:hmcode} shows residuals of matter power spectrum predictions from \texttt{HMCode2020} \cite{2021MNRAS.502.1401M} with respect to the test set drawn from the Mira-Titan CosmicEmu \cite{2017ApJ...847...50L}.
The \texttt{HMCode2020} predictions are as provided through \texttt{CAMB} 
\cite{2000ApJ...538..473L}.
Compared with Fig.~\ref{fig:matter_nmax3_cosmo_test_train}, the residuals are similar on the largest scales, slightly larger residuals on quasi-linear scales, and very slightly smaller on the smallest scales.
The residuals on quasi-linear scales appear larger than presented in Fig.~2 of \cite{2021MNRAS.502.1401M}, but are consistent with residuals with respect to Mira-Titan quoted in their Fig.~D1 (we have CosmicEmu in the numerator, so our residuals are inverted with respect to theirs).
The largest residuals at low $k$ trend with low-$\sigma_{8}$ models and the largest residuals at $0.1~\iM<k<1~\iM$ trend with both low-$\sigma_{8}$ models and high-$\Omega_{cb}$ models.
From this comparison, it is clear that the HZPT model is competitive when varying $\sigma_{8}$ and $\Omega_{cb}$.

\section{Halo correlator model components}
\label{appendix:halos}

\subsection{Model for $\xi_{hh}$ and $P_{hh}$ including halo exclusion}
\label{appendix:halos_exc}

Considering a general configuration-space model for the exclusion step $F_{\rm{exc}}(r)$, the two-point function of the discrete halo field is modeled as
\begin{equation}
    \xi_{hh}^{d}(r) -\frac{1}{\bar{n}}\delta^{D}(\mathbf{r})  = F_{\rm{exc}}(r) \left[1 + \xi_{hh}^{c}(r)\right] - 1
\label{eqn:2pt_excl_r}
\end{equation}
which gives in Fourier space
\begin{align}
    P_{hh}^{d}(k)
    = \frac{1}{\bar{n}} + \mathcal{F} \left( F_{\rm{exc}}(r) \left[1 + \xi_{hh}^{c}(r)\right] -1 \right)\\
     = \frac{1}{\bar{n}} + \left[\mathcal{F}\left( F_{\rm{exc}}(r) \right) * \mathcal{F} \left(1 + \xi_{hh}^{c}(r) \right)\right](k) - \delta^{(D)}(k) \\
     =\frac{1}{\bar{n}} + F_{\rm{exc}}(k) + \left[F_{\rm{exc}}(k) * P^{(c)}(k)\right],
\end{align}
where we have dropped the zero-lag Dirac delta in the final expression.

We may also write this expression in a slightly clearer way conceptually, conforming to the convention of B13 (see eqn.~\ref{eqn:2pt_excl_r}).
Instead of specializing to the too-simple top-hat model for the exclusion step (c.f. Figure~\ref{fig:excl_demo}), we assume only that we may write $F_{\rm{exc}}(r) = 1 - W(r)$, where $W(r)$ is a function with analytic Fourier transform.
Then, the expression is identical to eqn.~\ref{eqn:exclusion_P}, but we have taken the tiny step of generalizing the top-hat window to something more realistic:
\begin{equation}
    P_{hh}^{d}(k)
     =\frac{1}{\bar{n}} + P^{(c)}(k) - \tilde{W}(k) - \left[\tilde{W} * P^{(c)}\right](k).
\end{equation}
Also, we may define a finite-size stochastic contribution as the deviation of the true discrete power spectrum from the continuous model and the fiducial (Poisson) constant arising from discreteness:
\begin{align}
    S(k) = P^{(d)}(k)-\frac{1}{\bar{n}}-P^{(c)}(k) \\
          =  - \tilde{W}(k) - \left[\tilde{W}*P^{(c)}\right](k)
\end{align}
This is the quantity plotted in the right-hand panel of Figure~\ref{fig:excl_demo}, which quantifies the failure to include a model for exclusion in Fourier space.

As described in Section~\ref{section:halos}, we consider two effective models for the halo exclusion step, both of which perform much better than the simple top-hat.
The ErfLog model (almost the same as B13) is:
\begin{equation}
    F_{\mathrm{exc}}(r) = \frac{1}{2}\left[\mathrm{erf}\left(\frac{\log_{10}\left(\frac{r}{R_{\mathrm{exc}}}\right)}{\sigma_{\mathrm{exc}}}\right)+1\right]
\end{equation}

\noindent and the Exp model is 
\begin{equation}
    F_{\mathrm{exc}}(r,R_{\mathrm{exc}}) = \left[1-\exp\left(-\left(\frac{r}{R_{\mathrm{exc}}}\right)^{4} \right)\right]^{2}.
\end{equation}

The ErfLog model has no analytic Fourier transform, which must be computed numerically.
However, the Exp model has analytic Fourier transform, where the Fourier transform of the squared quantity is given by (defining the function $f$):
\begin{multline}
    \delta^{(D)}(k) \ + f(k,R_{\rm{exc}})=
    \delta^{(D)}(k) \ - \ \frac{R_{\rm{exc}}^{3}}{3}~\Gamma\left(\frac{7}{4}\right)~ {}_{0}F_{2}\left(\frac12,\frac54;\left(\frac{kR_{\rm{exc}}}{4}\right)^{4}\right) \\+ \ \frac{k^{2} R_{\rm{exc}}^{5}}{24}~ \Gamma\left(\frac{5}{4}\right)~ {}_{0}F_{2}\left(\frac32,\frac74;\left(\frac{kR_{\rm{exc}}}{4}\right)^{4}\right),
\end{multline}
where ${}_{0}F_{2}$ is the generalized hypergeometric function.
The full expression is then $F_{\mathrm{exc}}(k) = \delta^{(D)}(k) - 2 f(k,R_{\rm{exc}}) + f(k,2^{-\frac{1}{4}}R_{\rm{exc}})$.
It is helpful to consider this function from the perspective of modeling the nonperturbative exclusion effect in Fourier space (i.e. in the context of a Fourier space analysis), and in clearly disentangling the shape of the terms due to finite halo size and due to nonlinear clustering.
This form of the model is potentially computationally inefficient, but is presented for conceptual completeness, and convolutions may be sped up via FFTLog-based algorithms (e.g. \cite{2016JCAP...09..015M}).

\subsection{Halo mass bins}
\label{appendix:halos_bins}

\begin{figure}[h!]
\center
\includegraphics[width=5 in, angle=0]{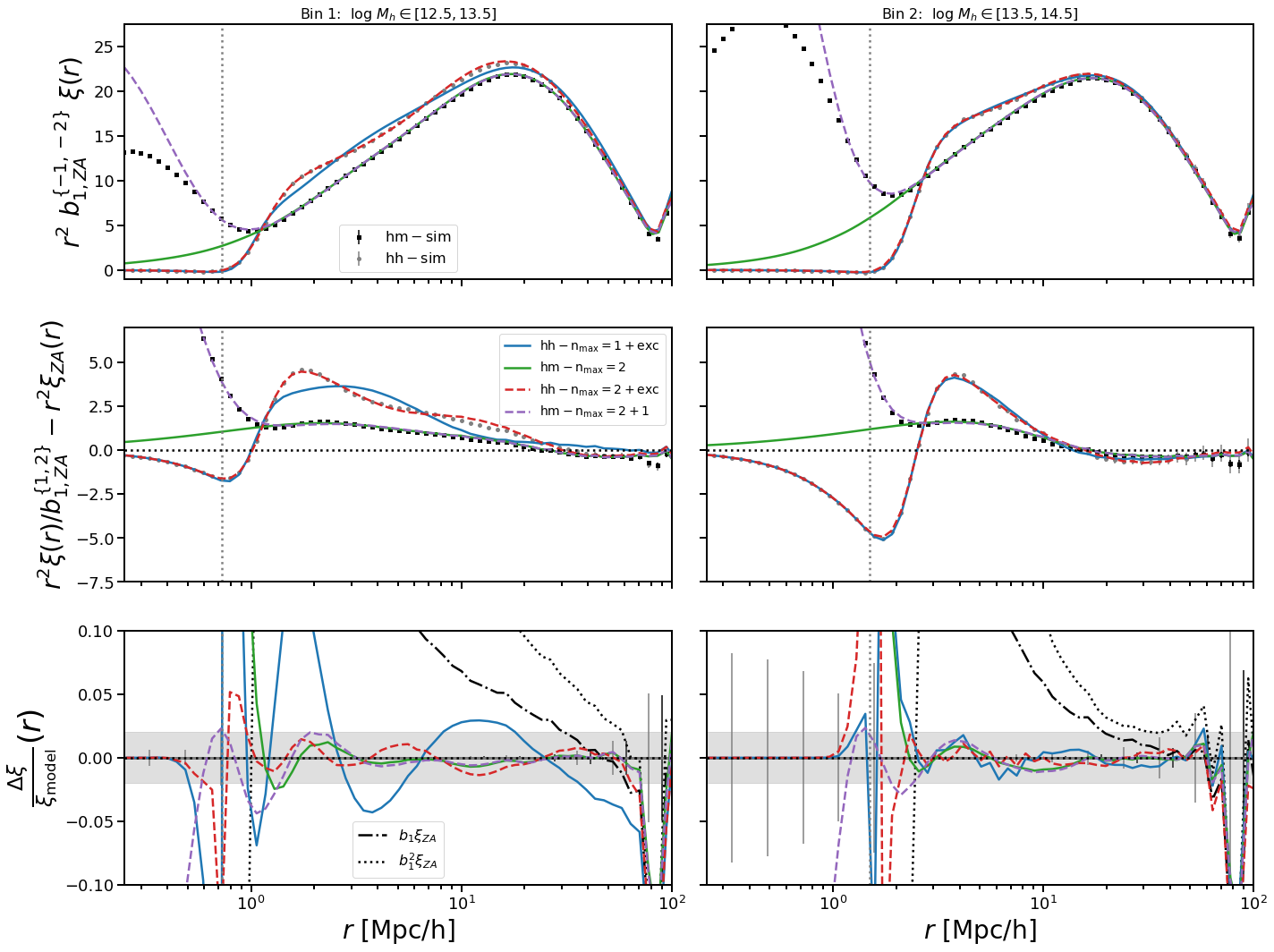}
\caption{Same as Fig.~\ref{fig:xih_resid}, but for the combined bins. 
\label{fig:halo_thick}}
\end{figure}

\begin{figure}[h!]
\center
\includegraphics[width=3 in, angle=0]{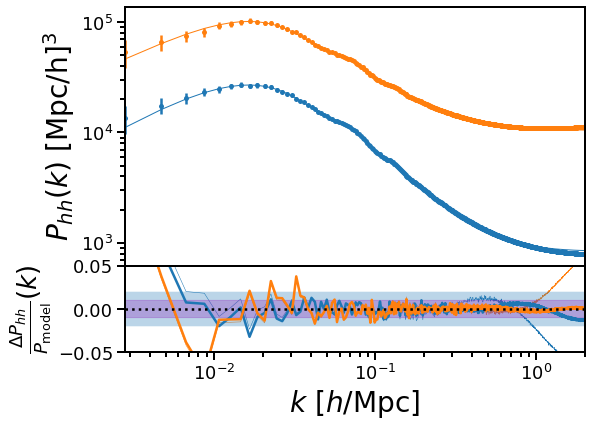}
\includegraphics[width=3 in, angle=0]{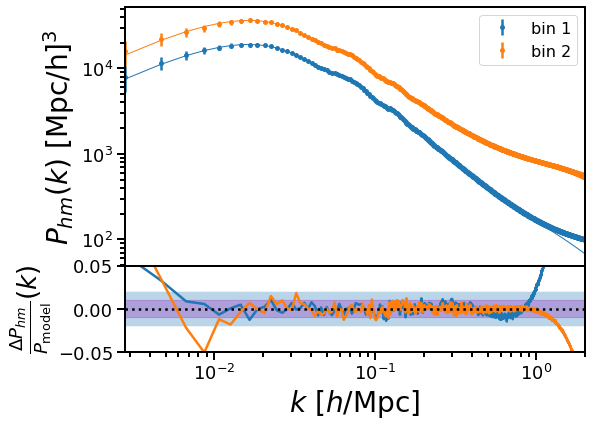}

\caption{Same as Fig.~\ref{fig:Ph_resid}, but for the combined bins.
\label{fig:Phalo_thick}}
\end{figure}

The halo bins used here are provided in Table~\ref{tab:halo_table}.
\begin{table}[h!]
    \centering
    \begin{tabular}{c|c}
        Bin & [$\log M_{\rm{min}},\log M_{\rm{max}}$]\\
        \hline
        1 & [12.5, 13.5]\\
        2 & [13.5, 14.5]\\
        3 & [11.5, 12.0]\\
        4 & [12.0, 12.5]\\
        5 & [12.5, 13.0]\\
        6 & [13.0, 13.5]\\
        7 & [13.5, 14.0]\\
        8 & [14.0, 14.5]\\
    \end{tabular}
    \caption{Halo mass bins in $M_{\odot} / h$}
    \label{tab:halo_table}
\end{table}
We do not show bins 0 and 3 in the main text since these low mass halos are almost certainly very poorly-resolved.
We also show the halo-matter and halo-halo correlation functions for wider bins that are the combination of the narrower two bins described in the main text (Fig.~\ref{fig:halo_thick}, Fig.~\ref{fig:Phalo_thick}).
The width of the bins seems not to adversely affect the accuracy of the model.
 

\subsection{Compensation $R$ for halos}
\label{appendix:halos_comp}

Fixing $R$ for halo auto-correlation to some very large value ($10^{10}$, essentially infinity for our purposes) has no real effect (fraction of 1\%) on the quality of fit in the halo auto-correlation for the $\nm=1$ model with free shot noise.
For halo-matter, the scale at which the compensation is relevant is different for different sized halos and increases with halo mass (we find this in the fitted values of $R$ for halo-matter).
Fixing $R$ to any particular value that works well for a certain mass bin produces significantly worse fits on large scales for other mass bins - so the compensation is not something that can easily be fixed for the cross correlation.
Disentangling compensation from nonlinear bias 
is also challenging since we expect both the physical compensation scale and nonlinear bias to change with halo mass.

\section{Impact of Halo Finder and Mass Definition on Halo Exclusion}
\label{appendix:halofinder}
The halo finder used has a significant impact on the scale and amplitude of the exclusion feature in the halo-halo correlation function.
The halo mass definition also has a moderate effect on the scale and significant impact on the amplitude of the feature.

\begin{figure}[h!]
\center
\includegraphics[width=4. in, angle=0]{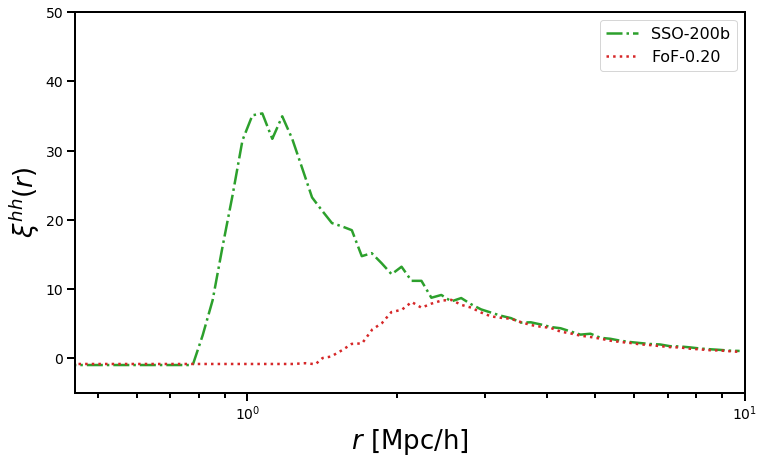}
\caption{Halo-halo correlation function for the default Aemulus ROCKSTAR “strict SO” (SSO) M200b halos (includes unbound particles) in green, and for FoF (linking length 0.2) halos in red.
\label{fig:halo_finder_ae}}
\end{figure}

\begin{figure}[h!]
\center
\includegraphics[width=4. in, angle=0]{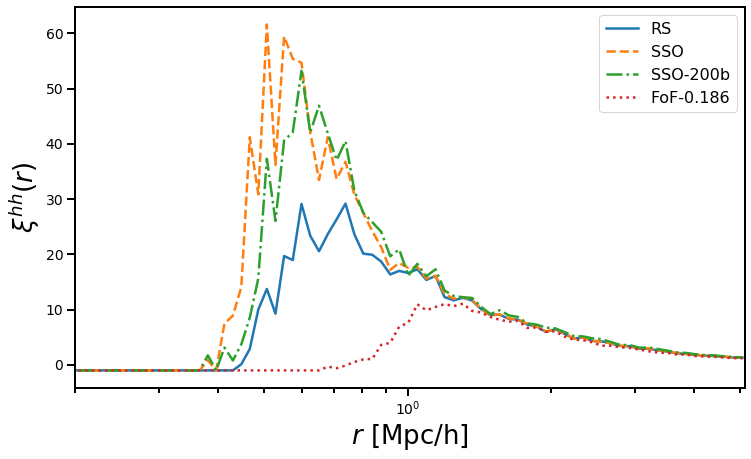}
\caption{Halo-halo correlation function for the Abacus \texttt{ROCKSTAR} default halos (blue), “strict SO” (SSO) virial halos (orange), SSO M200b halos (green), and for FoF with linking length 0.186 halos (red).
\label{fig:halo_finder_ab}}
\end{figure}

We saw in Section \ref{section:galaxies} that the exclusion feature presented itself at much smaller scales ($\approx 0.7 \ \M$) in the LZ correlation functions than in the CM correlation functions ($\approx 1.5 \ \M$).
To check if this effect is due to the choice of halo finder, we ran FoF on the Aemulus dark matter particle snapshots that were used to create the \texttt{ROCKSTAR} strict SO M200b halos provided in the Aemulus halo catalogs.
Fig.~\ref{fig:halo_finder_ae} shows the correlation function for both FoF halos and \texttt{ROCKSTAR} strict SO (SSO) halos.
It is clear that the correlation function of FoF halos exhibits an exclusion feature at much larger scales that the one for SSO halos does.
To check that this is not an effect unique to the Aemulus simulations, and to provide a direct comparison of SO and SSO mass definitions using the same halo centers, we use the publicly available Abacus halo catalogs \cite{2018ApJS..236...43G}.
The halo catalogs used correspond to Planck cosmology with box size 1100 $\M$ at $z=0.5$, and details of the FoF and \texttt{ROCKSTAR} catalogs are provided in \cite{2018ApJS..236...43G}.
Fig.~\ref{fig:halo_finder_ab} shows the same as Fig.~\ref{fig:halo_finder_ae} for the Abacus halos, but with additional curves for SO virial halos as well as default \texttt{ROCKSTAR} virial halos.
We consider only bin 7 here, but this effect persists for all halo mass bins considered in Table~\ref{tab:halo_table} (the only real differences being the absolute scales and smoother curves for lower-mass halos due to increased number density at lower mass).

It appears that for strict SO halos, halo mass definition does not change the exclusion feature significantly.
However, the use of default \texttt{ROCKSTAR} halo masses instead of SSO masses is already quite different in terms of the exclusion feature.
The exclusion features for the \texttt{ROCKSTAR} halo masses lie in the middle of those of the SSO and FoF halos.
The change in amplitude along with the change in scale is perhaps not that surprising, as it qualitatively seems to be what would be expected of applying the exclusion step to the continuous model for $\xi_{hh}(r)$ described in Section \ref{section:halos} at a smaller $R_{\rm{exc}}$.
Naively this would seem to mean that the SSO halos display clustering behavior indicative of a smaller exclusion scale (at fixed halo finder) than that exhibited by the \texttt{ROCKSTAR} halos (and the same goes for FoF halos).
This may be related to the well-known scatter in the relation between FoF ($b=0.2$) and M200b halos (e.g. \cite{2001A&A...367...27W},\cite{2009ApJ...692..217L}).
It would be interesting to connect this difference with a more intuitive description of exclusion, perhaps informed by peaks similar to \cite{2020arXiv201214404B}.
Figs. ~\ref{fig:halo_finder_ae} and ~\ref{fig:halo_finder_ab} show features that are similar to Fig. A1 of \cite{2019MNRAS.489.4170G} that compares different percolation strategies in a modified version of $\texttt{ROCKSTAR}$ (but did not investigate $\xi_{hh}(r)$ in detail). 
It would be interesting to draw clearer connections between the differences in percolation strategies, the differences we observe here, and the effect on halo exclusion in two-point correlators.

\section{Cross-correlation coefficient $r_{cc}$}
\label{appendix:rcc}

\begin{figure}[h!]
\center
\includegraphics[width=2.2 in, angle=0]{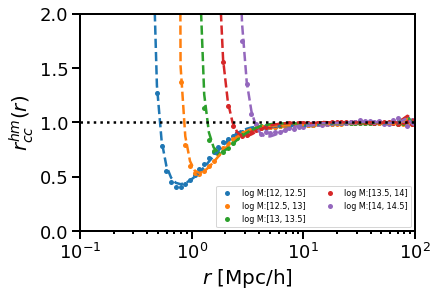}
\includegraphics[width=2.2 in, angle=0]{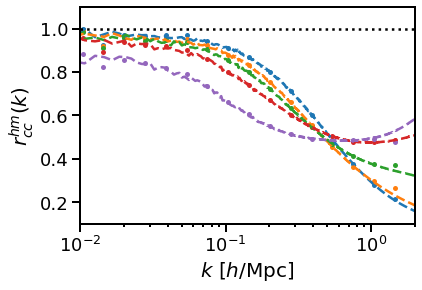}
\includegraphics[width=2.2 in, angle=0]{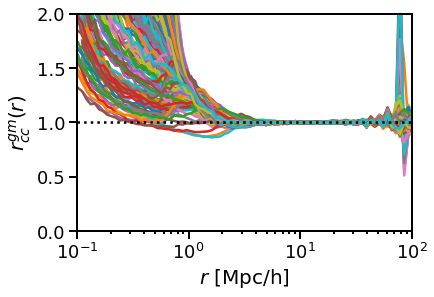}
\includegraphics[width=2.2 in, angle=0]{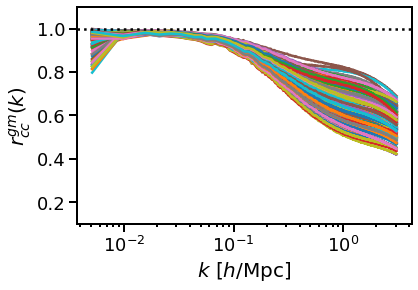}
\includegraphics[width=2.2 in, angle=0]{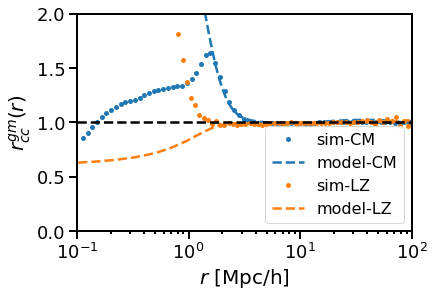}
\includegraphics[width=2.2 in, angle=0]{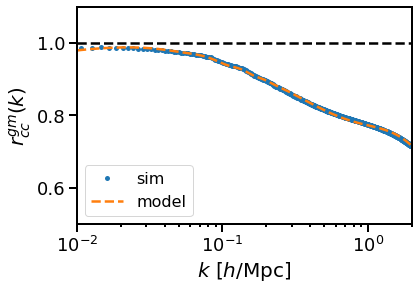}
\caption{
\textit{Top Left:} Cross-correlation coefficient for halos in configuration space.
\textit{Top Right:} Cross-correlation coefficient for halos in Fourier space. 
\textit{Middle Left:} Cross-correlation coefficient for (LOWZ) HOD galaxies in configuration space. 
\textit{Middle Right:} Cross-correlation coefficient for (LOWZ) HOD galaxies in Fourier space. 
Shot noise is \textit{not} subtracted for the Fourier space cross-correlation coefficients.
\textit{Bottom Left:} Cross-correlation coefficient and model for HOD galaxies in configuration space for single LOWZ and CMASS HOD realizations. 
\textit{Bottom Right:} Cross-correlation coefficient and model for HOD galaxies in configuration space for single CMASS HOD realization. 
Shot noise is \textit{not} subtracted for the Fourier space cross-correlation coefficients.
\label{fig:rcc}}
\end{figure}

The cross correlation coefficient provides a additional view of the non-perturbative effects present in small-scale two point statistics.
The cross correlation coefficient for two-point statistics is defined as $r^{xm}_{cc}(X) = \frac{X^{xm}}{\sqrt{X^{mm} X^{xx}}}$, where $x$ denotes the tracer type (galaxies or halos) and $X$ is the two-point statistic in question $P$ or $\xi$.
We present cross-correlation coefficients measured from the simulations and from the HZPT model for halos and galaxies in Figure~\ref{fig:rcc}.
The cross correlation coefficient for halos and galaxies is very close to 1 for all halo mass bins and HOD parameter choices in configuration space above $r \sim 4 \ \rm{Mpc}/h$.
On smaller scales, non-perturbative effects become obvious - the cross-correlation coefficient first drops in the region of the SSE and attains a minimum before blowing up as $r$ drops below $r_{\rm{exc}}$ at the exclusion step before becoming undefined as $\xi^{hh}(r)$ crosses zero.

For HOD galaxies, the picture is similar to the case of halos in a mass bin around $\log \frac{M}{M_{\odot}/h} \in [13 , 13.5]$, but with significant scatter due to the details of the halo occupation.
For many choices of halo occupation, this leads to $r_{cc}^{gm}(r)>1$ (but with a less severe growth toward small scales than for halos).
The undefined behavior at small scales for halos is also no longer present for galaxies since the satellite profile gives a non-zero contribution to $\xi_{gg}(r)$ on the scales at which $\xi_{hh}(r) \to -1$.
    
In Fourier space, the situation is perhaps more complicated due to finite halo size effects that manifest partially as sub/super-Poisson shot noise -  a more complete picture of finite-halo size effects is given by the modeling described in Section \ref{section:halos}.
By $k \sim 0.1 \ h/\rm{Mpc}$ $r_{cc}(k)$ already deviates from unity significantly for both halos and galaxies.
For halos, there is an offset from unity due to finite halo size (``sub/super-Poisson shot noise'' on large scales) that is more pronounced for larger halos, as expected.
Higher mass halos give $r_{cc}^{hm}(k)$ with a weaker scale dependence, leveling off quickly compared to lower mass halos, which continue to drop as $k$ increases.

\bibliographystyle{JHEP}
\bibliography{HZPT+.bib}

\end{document}